\documentclass[10pt]{article}

\usepackage{authblk}
\usepackage{graphicx}
\usepackage{geometry}
\usepackage[numbers]{natbib}
\expandafter\let\csname equation*\endcsname\relax
\expandafter\let\csname endequation*\endcsname\relax
\usepackage{amsmath,amsfonts,amssymb, mathtools}
\usepackage[dvipsnames]{xcolor}
\usepackage{braket}
\usepackage{physics}
\usepackage{subfig}
\usepackage{float}
\usepackage[pdftex]{hyperref}  
\hypersetup{
	colorlinks = true,
	linkcolor = blue,
	anchorcolor = blue,
	citecolor = blue,
	filecolor = blue,
	urlcolor = blue
    }
\usepackage{orcidlink}

\begin{document}
\title{Temporal evolution of a forced optomechanical system with linear and quadratic field — mechanical oscillator couplings}
\author[]{Luis A. Medina-Dozal\orcidlink{0000-0002-4695-5190}, Alejandro R. Urz\'ua\orcidlink{0000-0002-6255-5453}\thanks{Corresponding author: \href{mailto:alejandro@icf.unam.mx}{alejandro@icf.unam.mx}}, Jos\'e R\'ecamier-Angelini\orcidlink{0000-0002-5995-0380}}
\affil{Instituto de Ciencias F\'isicas, Universidad Nacional Autónoma de M\'exico\\ Avenida Universidad s/n, Col. Chamilpa, Cuernavaca, Morelos, 62210 Mexico.}
\maketitle


\begin{abstract}
In this work, we make use of Lie algebraic methods to obtain the time evolution operator for an optomechanical system with linear and quadratic couplings between the field and the mechanical oscillator. Firstly, we consider the case of a  non-driven system and find its exact time evolution operator, secondly we consider the case of a forced  system whose time evolution operator is obtained in an  approximate form. We confront our analytical results with a numerical simulation and find a good agreement between them.
\end{abstract}


\section{Introduction}
Quantum optomechanics explores the interaction between electromagnetic radiation and a nanomechanic or micromechanical motion. Since light carries linear and angular momentum, it gives rise to radiation pressure forces. These were postulated by Kepler and much later by Maxwell; the first demonstration of their existence is due to Levedew in 1901. Einstein \cite{Einstein:1909} derived the statistical properties of the radiation pressure fluctuations acting upon a movable mirror. In the 70s D. Wineland, C. Cohen-Tannoudji, W. D. Phillips, S. Chu, and  many others used resonant light scattering to achieve laser cooling of ions and of neutral atoms \cite{ Metcalf1999, Wineland75, Anderson1995} leading to the generation of Schr\"odinger cats \cite{Monroe1996} and atomic Bose-Einstein condensates \cite{Anderson1995,Ketterle1995}.  A.  Ashkin demonstrated that small dielectric balls can be manipulated using radiation pressure forces associated with focused laser beams \cite{Ashkin78} leading to the realization of optical tweezers.  

Non-resonant light-matter interactions have the advantage of being wavelength-independent, thus allowing the achievement of optomechanical effects in a wide range of wavelengths from microwave to the optical domain. Resonant interactions have a much more limited range of wavelengths but a much larger intensity. 

In the 90s several aspects of quantum cavity-optomechanical systems started to be explored theoretically, among them the squeezing of light \cite{Mancini1994},  and Quantum Non-Demolition detection of the light intensity which makes use of the Kerr-like nonlinearity introduced by the optomechanical interaction \cite{Jacobs1994} as well as the nonclassical states of the field and the mechanical oscillator when the optomechanical coupling is large enough \cite{Knight97,Manko97}. 

The simplest optomechanical system in the optical domain consists of a laser driven optical cavity with  a vibrating end mirror. However, optomechanical coupling has been reported in many systems,  for instance in membranes and nanorods inside a Fabry-P\'erot cavity \cite{Thompson2008,Favero2009}.
Reference  \cite{Aspelmeyer14} is an excellent review dealing with optomechanical systems dated up to 2014.

 Rai and Agarwal \cite{Rai2008} showed that a \emph{quantum optical spring}, undergoes revivals in their dynamical observables when the radiation pressure of a field is coupled to the quadratic quadrature of a mechanical oscillator. Under the selection of coherent states as initial conditions, the evolution profile depends on the sum of the field modes; no linear term of interaction was considered, it was a purely quadratic Hamiltonian. Later, Shi and Bhattacharya \cite{Shi2013} give a more complete survey on this system. In this matter, studies exist on how the purely quadratic coupling Hamiltonian can be used in optomechanics to achieve phonon cooling and squeezing \cite{Nunnenkamp2010}. Liao and Nori used this quadratic platform to study single photon coherent interaction in a cavity with a membrane inside\cite{jie14}. Photon blockade \cite{Xie2016} and circuit analog \cite{Kim2015} were shown. Cross-correlations between photons and phonons were studied in \cite{Xu2018}. Furthermore, the quadratic coupling can be enhanced by employing a nonlinear medium and lasers \cite{Zhang2019}. Quantum signatures in quadratic optomechanics  have also been explored \cite{machado2019}, showing that the nonlinear term enables the  observation of effects similar to those in the single photon - strong coupling regime. Recently, a hybrid quadratic optomechanical system was used to study its optical response and coherence \cite{Kundu2021}. On the other side, the modeling of linear and quadratic couplings in optomechanical systems can be achieved selectively via conditional measurements \cite{Vanner2011}. The effects of the dispersive regime in linear and quadratic couplings found an explanation using optical squeezing, showing how the presence or absence of one of the coupling terms affects the stability and optical quadrature of the system \cite{Satya2019}. There are accounts studying interactions with higher orders in quantum optomechanics \cite{Khorasani2017}. Induced transparency (the ability of a medium to be transparent, or translucent, in a frequency spectrum narrowband) can be achieved using both couplings in the optomechanical setup \cite{Zhang2018}. Simultaneous optical and mechanical squeezing was found in a configuration where two cavities sharing a mechanical resonator experience linear and quadratic couplings \cite{Gu2019}. Using a similar platform, photon hopping enables the coexistence of linear and quadratic couplings under the rotating wave approximation, enabling the proposal of a dual-mode sensor in an optomechanical system \cite{Chao2021}.


\section{Theory}
The simplest optomechanical system is composed of a Fabry-Perot  resonator with a movable mirror, which couples with a mechanical oscillator. These resonators can be realized experimentally in different forms, and it is then possible to attain a wide range of coupling constants. Because of the radiation pressure, the length of the cavity is modified and, as a consequence, the resonance frequency  becomes a function of the mirror's position \cite{kippenberg}. The Hamiltonian describing this system can be written as  \cite{Meystre2013}: 
\begin{equation}
    \hat H = \hbar \omega(x) \hat n +\hbar \omega_m \hat N
\end{equation}
with $\hat{N}= \hat{b}^{\dagger}\hat{b}$ the phonon number operator, and $\hat{n} = \hat{a}^{\dagger}\hat{a}$ the photon number operator. The resonance frequency, assuming small displacements from the original position, $\frac{\hat{x}}{L}\ll1 $ is: 
\begin{equation}
    \omega(x) =\frac{\pi c}{L+\hat x} = \omega_c\left(\frac{1}{1+\frac{\hat x}{L}}\right)\simeq \omega_c(1-\frac{\hat x}{L}+\frac{\hat x^2}{L^2}),
\end{equation}
with $\{\omega_c=\frac{\pi c}{L},L\}$, the frequency and the initial length of the cavity. 
Now we define the displacement from the equilibrium position as
$\hat x = \sqrt{\frac{\hbar}{2m\omega_m}}(\hat b +\hat b^{\dagger}) = x_{zpt}(\hat b+\hat b^{\dagger})$
where the operators $\hat{b}$, $\hat{b}^{\dagger}$ are the annihilation and creation operators for the mechanical oscillator and satisfy the commutation relation $[\hat{b},\hat{b}^{\dagger}]= 1$. \\
The Hamiltonian is then given as
\begin{equation}\label{eq:Ham}
    \frac{\hat H}{\hbar} = (\omega_c+G_1)\hat n+ \omega_m \hat N -G_0\hat n(\hat b+\hat b^{\dagger})+2G_1\hat n\hat N+G_1\hat n(\hat b^2+\hat b^{\dagger 2}).
\end{equation}
where we have defined $G_0/\omega_c= x_{zpt}/L$, $G_1 = G_0^2/\omega_c$ and these constants have units of frequency. As mentioned above, in the general case,  these couplings can be considered as independent since their values depend upon the specific experimental realization \cite{purdy10}. 


\subsection{Non forced optomechanical system with linear coupling}\label{sec:linearcoupling}
We start with the simple case of a linear coupling and an unforced system \cite{Knight97, Manko97, recamier24} and assume that the retardation effects, due to the oscillating mirror, are negligible. We also assume that the mechanical frequency $\omega_m$ is orders of magnitude smaller than the resonant frequency of the cavity, this condition implies that the photons generated by the Dynamical Casimir Effect (DCE) in the resonator are negligible \cite{roman17}. The Hamiltonian is then given by:
\begin{equation}\label{eq:Hamlin}
    \frac{\hat H_{opt}}{\hbar} = \omega_c\hat n+ \omega_m \hat N -g_0\omega_m\hat n(\hat b+\hat b^{\dagger}).
\end{equation}
where we have written the coupling in terms of the oscillator's frequency $G_0 = g_0 \omega_m$. 
In  the following table, we show the commutation relations for the operators appearing in Eq.~\ref{eq:Hamlin}.

\begin{center}
\begin{tabular}{|c|c|c|c|c|c|}
\hline
&$\hat{n}$&$\hat{N}$&$\hat{n}\hat{b}$&$\hat{n}\hat{b}^{\dagger}$&$\hat{n}^2$\\
\hline
$\hat{n}$&0&0&0&0&0\\
\hline
$\hat{N}$&0&0&$-\hat{n}\hat{b}$&$\hat{n}\hat{b}^{\dagger}$&0\\
\hline
$\hat{n}\hat{b}$&0&$\hat{n}\hat{b}$&0&$\hat{n}^2$&0\\
\hline
$\hat{n}\hat{b}^{\dagger}$&0&$-\hat{n}\hat{b}^{\dagger}$&$-\hat{n}^2$&0&0\\
\hline
$\hat{n}^2$&0&0&0&0&0 \\
\hline
\end{tabular}
\end{center}
Notice that we have to incorporate the operator $\hat n^2$ to have a closed Lie algebra, thus the optomechanical interaction induces a Kerr-like nonlinearity. The optomechanical Hamiltonian can then be written as:
\begin{equation}
\hat{H}_{opt} = \sum_{n=1}^{5} \Phi_{n}\hat{X}_n 
\end{equation}
with $\hat{X}_1=\hat{n}$, \ $\hat{X}_2 = \hat{N}$, \  $\hat{X}_3 = \hat{n}\hat{b^{\dagger}}$, \  $\hat{X}_4 = \hat{n}\hat{b}$ and $\hat{X}_5 = \hat{n}^2$.

Using the Wei-Norman theorem \cite{Manko97, wei-norman, recamier2011, recamier-entropy} we can write the {\em exact} time evolution operator as,
\begin{equation}\label{eq:Uopt}
\hat{U}_{opt} = \prod_{n=1}^{5} e^{\alpha_n(t) \hat{X}_n} = e^{\alpha_1(t) \hat{n}} e^{\alpha_2(t)\hat{N}} e^{\alpha_3(t) \hat{n}\hat{b^{\dagger}}} e^{\alpha_4(t)\hat{n}\hat{b}} e^{\alpha_5(t)\hat{n}^2}
\end{equation}
In this simple case, it is possible to obtain the explicit form for the functions $\alpha_i$. We find that  $\alpha_3 (t)=-\alpha_4^{*}(t)$ and then the time evolution operator can be written as:
 \begin{equation}
 \hat{U}_{opt} = e^{\alpha_1(t) \hat{n}} e^{\alpha_2(t)\hat{N}} e^{(\frac{1}{2}|\alpha_3(t)|^2+\alpha_5(t))\hat{n}^2}\hat{D}_{\hat{b}}(\alpha_3(t)\hat{n}).
 \end{equation} 
 with $\hat{D}_{\hat{b}}(\alpha_3(t)\hat{n})$ Glauber's displacement operator,
 \begin{equation}
      \hat{D}_{\hat{b}}(\alpha_3(t)\hat{n}) = e^{\alpha_3(t) \hat{n}\hat{b}^{\dagger} -\alpha_3^{*}(t)\hat{n}\hat{b}}.
 \end{equation}
 
 Usually, the coupling constant is such that $g_0 = G_0/\omega_m\ll 1$. \cite{verhagen2012}.

\begin{figure}[hbtp]
    \centering
    \includegraphics[width = \linewidth]{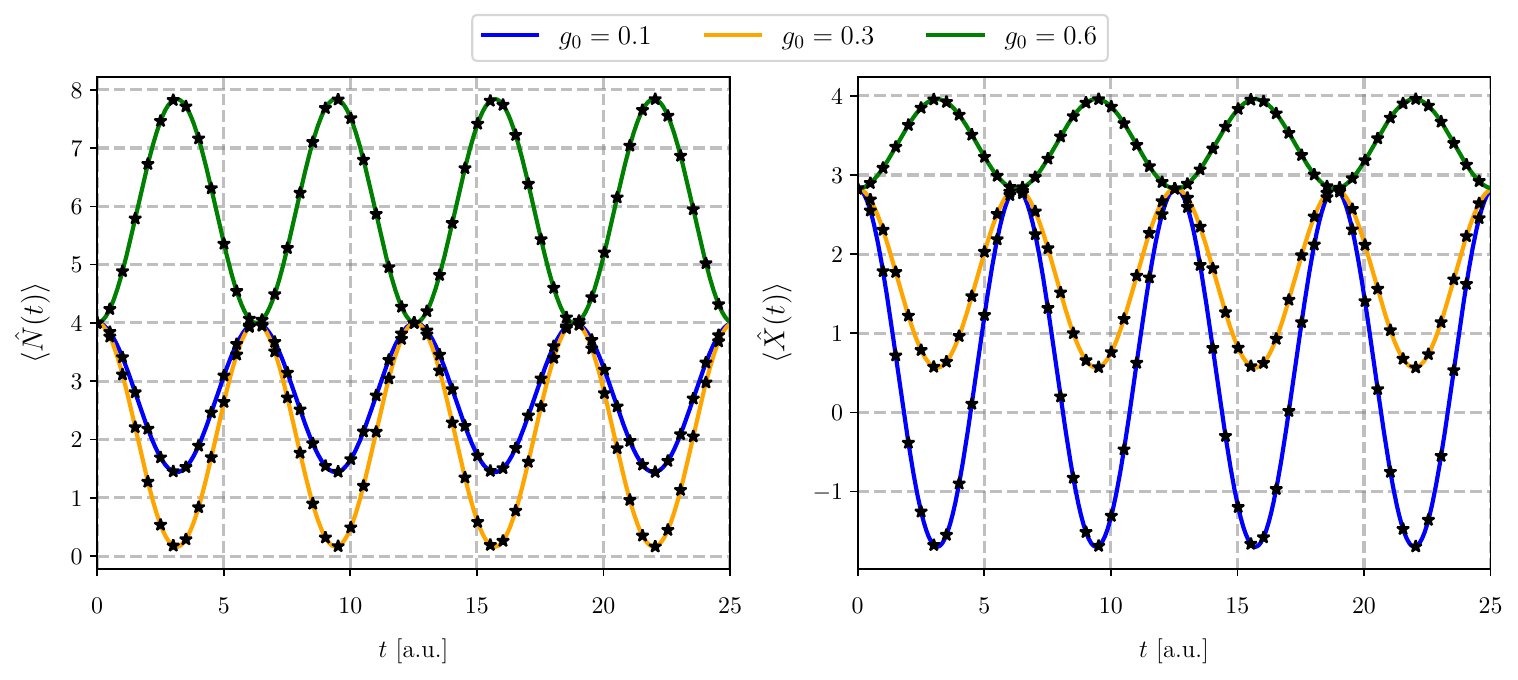}
    \caption{Average value of the phonon number operator (left panel) and the position quadrature (right panel) for an initial state with $n=4$ photons in the cavity and a mechanical coherent state with $\Gamma=2$. The coupling constants are $g_0=0.1$ (blue), $g_0=0.3$ (orange) and $g_0=0.6$ (green). The oscillator's frequency is $\omega_m=1$. Black star marks denote a simulation running, having one-one correspondence with the analytical result.}
    \label{Naverage}
\end{figure}

 Once we know the time-evolution operator, we can evaluate the average value of any observable. In figure \ref{Naverage} we show the average value of the phonon number operator for an initial state $|\Psi(0)\rangle = |n\rangle \otimes |\Gamma\rangle$ with $|n\rangle$ a photon number state and $|\Gamma\rangle$ a mechanical coherent state. In this case, the average is given by,
 \begin{equation}
     \langle \Psi(t)|\hat{N}|\Psi(t)\rangle = |\Gamma e^{-\mathrm{i}\omega_m t}+ g_0 (1-e^{-\mathrm{i}\omega_m t})n|^2
 \end{equation}
 where we have substituted the explicit form for the coefficient $\alpha_3(t)$. One can notice the evident entanglement between the field and the mechanical oscillator in the dependence of the mechanical coherent state upon the number of photons present in the cavity.

In figure \ref{Naverage} we show the temporal evolution of the average phonon number operator for several values of the coupling constant $G_0$. We see that, for small values of the coupling constant, the phonon number operator is a periodic decreasing function of time starting at $\langle N\rangle=4$ and attaining values of the order of $\langle N \rangle \simeq 2$. The frequency of the oscillations is equal to that of the mechanical oscillator $\omega_m$.  When we increase the coupling to $G_0=0.3 \omega_m$,  the average falls to almost zero phonons at the minimum. When we increase even more the coupling to $G_0=0.6 \omega_m$, the mechanical oscillator modifies its conduct drastically and increases the average number of phonons attaining values up to $N=8$. This result was reported in \cite{recamier24} 


\subsection{Non forced optomechanical system with quadratic coupling}\label{sec:quadratic}

In some optomechanical systems, an optical cavity mode is coupled to the square of the position of the mechanical oscillator. An example is the membrane-in-the-middle geometry, where the membrane is placed in a node or an antinode of the cavity field \cite{Thompson2008, jayich08, sankey10, girvin10}. 
In the Hamiltonian given by Eq.~\ref{eq:Ham} we now set $G_0=0$ and set $G_1=g_1 \omega_m$ and retain the quadratic coupling. Then we get:
\begin{equation}
    \frac{\hat{H}}{\hbar} = (\omega_c+g_1\omega_m)\hat{n} + \omega_m \hat{N} + 2g_1\omega_m \hat{n}\hat{N}+ g_1\omega_m (\hat{b}^2 +\hat{b}^{\dagger 2})
\end{equation}
Notice that the photon number operator is a constant of the motion since it commutes with the Hamiltonian, then, it can be interpreted as a simple parameter specifying the initial number of photons in the cavity. The Hamiltonian is,
\begin{equation}\label{eq:Hcuad}
    \frac{\hat{H}^{(n)}}{\hbar} = (\omega_c+g_1\omega_m)n + \omega_m(1+2g_1 n) \hat{N} + g_1\omega_m(\hat{b}^2 +\hat{b}^{\dagger 2}) \equiv \sum_{k=1}^4 \Phi_k^{(n)} \hat{X}_k.
\end{equation}
The set of operators  appearing in the Hamiltonian $\{ \hat{N}, \hat{b}^2, \hat{b}^{\dagger 2}, {\cal{I}} \}$ has the commutation relations given in the following table:
\begin{center}
\begin{tabular}{|c|c|c|c|}
\hline
&$\hat{b}^2$&$\hat{b}^{\dagger 2}$&$\hat{N}$\\
\hline
$\hat{b}^2$&0&$4\hat{N}+2$&$2\hat{b}^2$\\
\hline
$\hat{b}^{\dagger 2}$&$-4\hat{N}-2$&0&$-2\hat{b}^{\dagger 2}$\\
\hline
$\hat{N}$&$-2\hat{b}^2$&$2\hat{b}^2$&0 \\
\hline
\end{tabular}
\end{center}
and we see that the Hamiltonian forms a finite Lie algebra. Then, we can write the {\em exact} time evolution operator as
\begin{equation}
    \hat{U}^{(n)}_{q}(t)= \prod_{k=1}^{4} e^{\beta^{(n)}_k(t) \hat{X}_k}
\end{equation}
with $\hat{X}_1=\hat{b}^{\dagger 2}$, $\hat{X}_2 = \hat{N}$, $\hat{X_3} = \hat{b}^2$ and the functions $\beta_i^{(n)}(t)$ such that:
\begin{equation}
\begin{aligned}
    \dot{\beta}_1^{(n)}(t) & =  -\mathrm{i}\left(\Phi_1^{(n)}+4\beta_1^{(n){2}}(t)\Phi_3^{(n)} +2\beta_1^{(n)}(t) \Phi_2^{(n)} \right) \\
    \dot{\beta}_2^{(n)}(t) & =  -\mathrm{i}\left(\Phi_2^{(n)} + 4\beta_1^{(n)}(t) \Phi_3^{(n)}\right) \\
    \dot{\beta}_3^{(n)}(t) & =  -\mathrm{i}\, \Phi_3^{(n)} e^{2\beta_2^{(n)}(t)} \\
    \dot{\beta}_4^{(n)}(t) & =  -\mathrm{i}\left( \Phi_4^{(n)} + 2\beta_1^{(n)}(t) \Phi_3^{(n)}\right)
\end{aligned}
\end{equation}
with the initial conditions $\beta_i^{(n)}(0)=0.$ These equations are solved with Mathematica.

Since we have found an explicit expression for the time evolution operator, we can transform any operator to the Heisenberg representation.
The operators $\hat{b}$, $\hat{b}^{\dagger}$ in the Heisenberg representation are:
\begin{equation}\label{eq:bheis}
\hat{b}^{\dagger}(t) = e^{-\beta_2^{(n)}(t)}\left(\hat{b}^{\dagger}-2\beta_3^{(n)}(t) \hat{b} \right), \ \hat{b}(t) = \hat{b}\left(e^{\beta_2^{(n)}(t)}-4\beta_1^{(n)}\beta_3^{(n)} e^{-\beta_2^{(n)}}\right) +2\beta_1^{(n)}e^{-\beta_2^{(n)}}\hat{b}^{\dagger}
\end{equation} 
The phonon number operator in the Heisenberg representation is:
\begin{equation}
\begin{aligned}
    \hat{N}(t) & =   \hat{b}^{\dagger}(t)\hat{b}(t)  =  \left(1-8\beta_1^{(n)}(t)\beta_3^{(n)}(t)e^{-2\beta_2^{(n)}(t)}\right)\hat{N} -4\beta_1^{(n)}(t)\beta_3^{(n)}(t)e^{-2\beta_2^{(n)}(t)}{\cal{I}} \\
    &+ 2\beta_1^{(n)}(t)e^{-2\beta_2^{(n)}(t)}\hat{b}^{\dagger 2} + 2\beta_3^{(n)}(t)\left(4\beta_1^{(n)}(t)\beta_3^{(n)}(t)e^{-2\beta_2^{(n)}(t)}-1 \right) \hat{b}^2 
\end{aligned}
\end{equation}
its average value for an initial state $|\Psi(0)\rangle = |n\rangle\otimes |\Gamma\rangle$ is:
\begin{equation}
\begin{aligned}
    \langle n, \Gamma|\hat{N}(t)| n,\Gamma\rangle & =  \left(1-8\beta_1^{(n)}(t)\beta_3^{(n)}(t)e^{-2\beta_2^{(n)}(t)}\right) |\Gamma|^2 - 4\beta_1^{(n)}(t)\beta_3^{(n)}(t)e^{-2\beta_2^{(n)}(t)}  \\
    & + 2\beta_1^{(n)}(t)e^{-2\beta_2^{(n)}(t)} \Gamma^{* 2} + 2\beta_3^{(n)}(t)\left(4\beta_1^{(n)}(t)\beta_3^{(n)}(t)e^{-2\beta_2^{(n)}(t)}-1 \right) \Gamma^2
\end{aligned}
\end{equation}
and for an initial state, $|\Psi(0)\rangle = |\alpha\rangle\otimes |\Gamma\rangle $ it is:
\begin{equation}
\langle \alpha,\Gamma|\hat{N}(t)|\alpha,\Gamma\rangle = e^{-|\alpha|^2} \sum_{k=0}^{\infty} \frac{|\alpha|^{2k}}{k!}\langle k,\Gamma|\hat{N}(t)|k,\Gamma\rangle 
\end{equation}
The quadratures
\begin{equation}
    \hat{X}(t) = \frac{1}{\sqrt{2}} \left(\hat{b}(t)+\hat{b}^{\dagger}(t)\right), \ \hat{P}(t) = \frac{\mathrm{i}}{\sqrt{2}}\left(\hat{b}^{\dagger}(t) - \hat{b}(t) \right) 
\end{equation}
and their dispersions are obtained from Eq.\ref{eq:bheis}.
\begin{figure}[H]
    \centering
    \includegraphics[width = \linewidth]{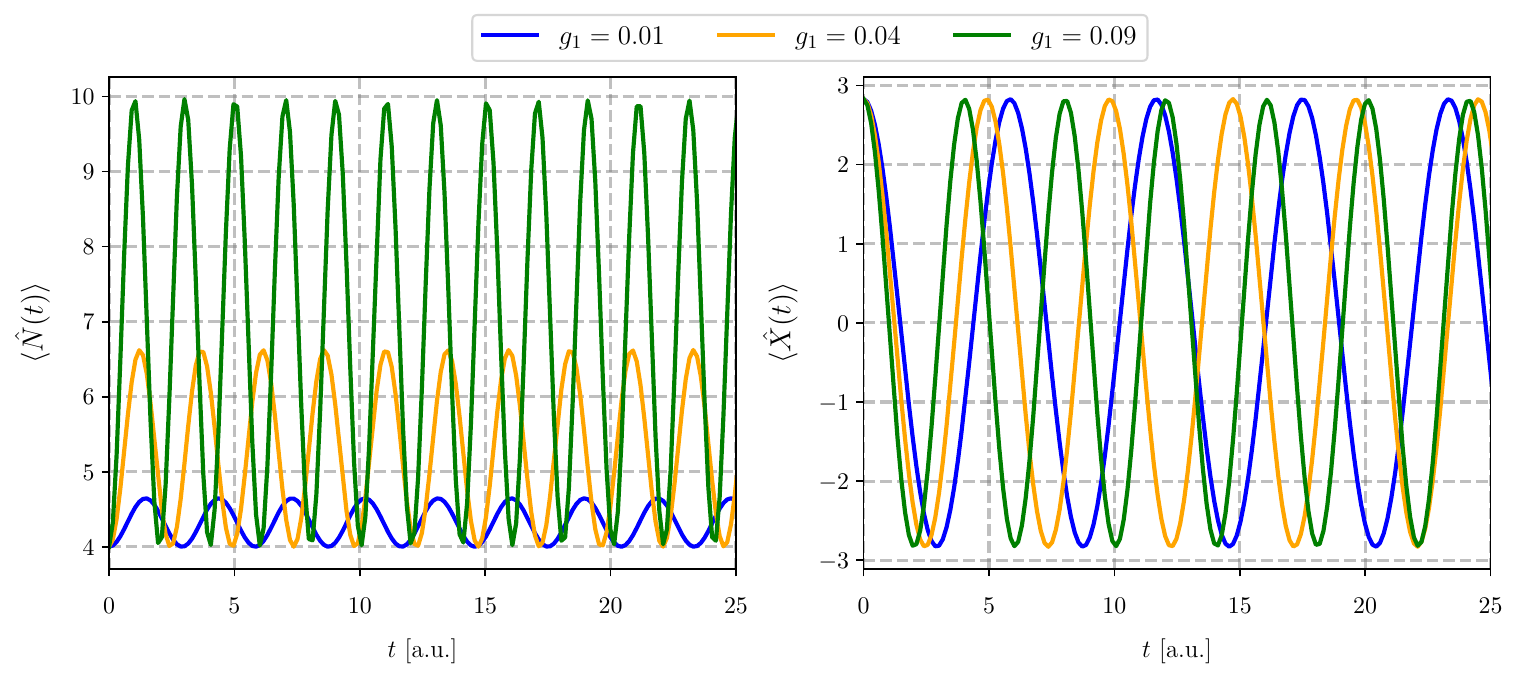}
    \caption{Left panel, average value of the phonon number operator for an initial state $|n\rangle |\Gamma\rangle$ with $n$ a number state for the field and $|\Gamma\rangle$ a coherent state for the mechanical oscillator. Right panel, average value of the position quadrature $\hat{X}(t)$.  Here we used $n=4$, $\Gamma=2$, $g_0=0.0$, $g_1=0.01$ (blue), $g_1=0.04$ (orange) and $g_1=0.09$ (green). The mechanical oscillator's frequency is $\omega_m = 1$.}
    \label{Navercuad}
\end{figure}

In figure \ref{Navercuad} left panel, we show the temporal evolution of the average value of the phonon number operator and in the right panel the evolution of the position quadrature, both for an initial state $|n,\Gamma\rangle$ with $n=4$, $\Gamma=2$ and $G_0=0$, $g_1=0.01$ (blue), $g_1=0.04$ (orange) and $g_1=0.09$ (green). We can see that the average number of phonons is an increasing periodic function of time, with a frequency and amplitude proportional to the coupling intensity. The frequency of the oscillations is, $\omega= 2 \omega_m^{(eff)}= 2 \omega_m (1+2g_1 n)$ since in this case, we only have quadratic terms in the creation-annihilation operators for the mechanical oscillator.  Regarding the average position, we see that the amplitude of the oscillations is the same regardless of the intensity of the quadratic coupling, although the frequency is $\omega= \omega_m(1+2g_1 n)$ and is different for each value of the coupling parameter. 
\begin{figure}[hbtp]
    \centering
    \includegraphics[width = \linewidth]{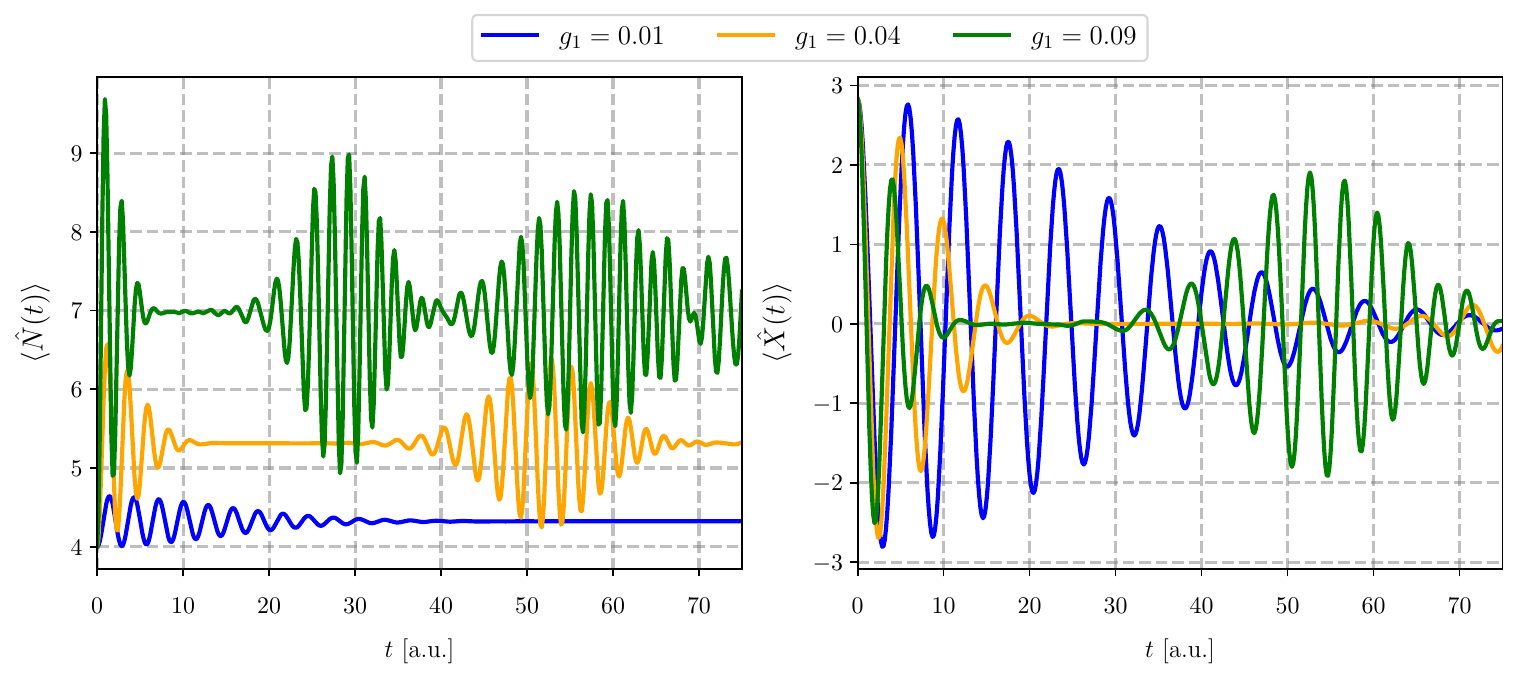}
    \caption{Left panel: average value of the phonon number operator for an initial field coherent state with $\alpha=2$  and a mechanical coherent state with $\Gamma=2$. Right panel: average value of the quadrature $\hat{X}(t)$ with the same initial state. The coupling constants are $g_0=0.0$, $g_1=0.01$ (blue), $g_1=0.04$ (orange) and $g_1=0.09$ (green).}
    \label{Navercoh}
\end{figure}

In figure \ref{Navercoh} we show the temporal evolution of the average phonon number operator (left panel) and the average quadrature $\bar{X} =\langle \hat{X}(t)\rangle$ (right panel) for an initial state $|\alpha,\Gamma\rangle = |2,2\rangle$ with $\alpha$ a coherent state for the field and $\Gamma$ a coherent state for the mechanical oscillator. The coupling parameters are stated in the caption. At the initial time, the average phonon number operator is $N=4$ as corresponds to the coherent state with $\Gamma=2$. As time evolves, it oscillates above the initial value, and after some oscillations it attains a collapse similar in form to those present in the Jaynes-Cummings model. This conduct is maintained along the evolution, with revivals with longer duration and smaller intensity. Notice also that the appearance of the revivals is faster as the intensity of the coupling increases.  
We see in the right panel that the position quadrature shows oscillations around the origin. At the beginning of the evolution, the oscillations have the largest amplitude, similar to what we found for an initial state with a fixed number of photons (see Fig.\ref{Navercuad}). After some time, we see the appearance of a collapse and thereafter, a revival of the oscillations. The appearance of the collapses is faster as the intensity of the coupling increases. The timescale in both figures is the same, so that collapses and revivals are more frequent in the evolution of the phonon number operator. This conduct was also reported in \cite{machado2019}, where they studied a purely quadratic interaction. When we have only the linear coupling, these collapses and revivals are not present even with a field coherent state as the initial state.


\subsection{Non forced optomechanical system with linear and quadratic couplings}\label{sec:linplusquad}
Now we retain all the terms in the Hamiltonian \ref{eq:Ham}
\begin{equation}
    \frac{\hat H}{\hbar} = (\omega_c+g_1\omega_m)\hat n + \omega_m \hat N -g_0\omega_m\hat n(\hat b+\hat b^{\dagger})+2g_1\omega_m\hat n\hat N+g_1\omega_m\hat n(\hat b^2+\hat b^{\dagger 2}).
\end{equation}
Notice, that the photon number operator is a constant of the motion so that, as we did before, it may be considered a parameter defining a given Hamiltonian $\hat{H}^{(n)}$. With this in mind, we write the Hamiltonian, for a given $n$ as:
\begin{equation}\label{eq:Hnf}
    \frac{\hat{H}^{(n)}}{\hbar} = (\omega_c+g_1\omega_m) n +\omega_m(1 +2g_1 n)\hat{N} - g_0\omega_m(\hat{b}+\hat{b}^{\dagger})+ g_1\omega_m n(\hat{b}^2+\hat{b}^{\dagger 2})  
\end{equation}
and we see that the Hamiltonian contains only operators corresponding to the Mechanical oscillator. The set $\{ \hat{b}, \hat{b}^{\dagger}, \hat{N}, \hat{b}^2, \hat{b}^{\dagger 2}, {\cal I} \}$ is closed under commutation, and we can write the time-evolution operator as
\begin{equation}\label{eq:Unl}
    \hat{U}_{nl}^{(n)}(t) = \prod_{k=1}^6 e^{\alpha_k^{(n)}(t) \hat{X}_k}
\end{equation}
where $\hat{X}_1=\hat{b}^{\dagger 2}$, $\hat{X}_2 =\hat{b}^{\dagger}$, $\hat{X}_3 = \hat{N}$, $\hat{X}_4 = \hat{b}$, $\hat{X}_5 = \hat{b}^2$, $\hat{X}_6 = {\cal I} $, and
with complex, time-dependent functions $\alpha_k^{(n)}(t)$ to determine. Substitution of Eq.~\ref{eq:Unl} in Schr\"odinger's equation yields a set of six coupled, nonlinear, ordinary differential equations for the functions $\alpha_k^{(n)}(t)$. We solve this system of equations by numerical means. Once we know the explicit form of the time evolution operator, we can calculate the average value of any observable.

Let us consider the average value of the phonon number operator.
\begin{equation}
    \langle \hat{N}(t)\rangle = \langle \Psi(0)|\hat{U}_{nl}^{(n)\dagger}(t)\hat{N}\hat{U}_{nl}^{(n)}(t)|\Psi(0)\rangle
\end{equation}

Transforming the operators $\hat{b}^{\dagger}$, $\hat{b}$ yields
\begin{equation}\label{eq:transbd}
\begin{aligned}
\hat{b}^{\dagger}(t) = \hat{U}_{nl}^{(n)\dagger} \hat b^{\dagger} \hat{U}_{nl}^{(n)}  &=  e^{\mathrm{i}\omega_m t}\left(e^{-\alpha_3^{(n)}(t)}\hat{b}^{\dagger} -2\alpha_5^{(n)}(t)e^{-\alpha_3^{(n)}(t)}\hat{b} -\alpha_4^{(n)}(t)e^{-\alpha_3^{(n)}(t)} \right)\\
& \equiv f_1^{(n)}(t)\hat{b}^{\dagger} + f_2^{(n)}(t)\hat{b} + f_3^{(n)}(t)
\end{aligned}
\end{equation}
\begin{equation}\label{eq:transb}
\begin{aligned}
\hat{b}(t) =  \hat U^{(n)\dagger}_{nl}\hat{b}\hat U^{(n)}_{nl} & =  e^{-\mathrm{i}\omega_m t}\left( \left(e^{\alpha^{(n)}_3(t)}-4\alpha^{(n)}_1(t)\alpha^{(n)}_5(t) e^{-\alpha^{(n)}_3(t)} \right)\hat b +2\alpha^{(n)}_1(t) e^{-\alpha^{(n)}_3(t)}\hat b^{\dagger}\right)\\ & +  e^{-\mathrm{i}\omega_m t}\left( \alpha^{(n)}_2(t) - 2\alpha^{(n)}_1(t)\alpha^{(n)}_4(t)e^{-\alpha^{(n)}_3(t)} \right) \\ &  \equiv  f_4^{(n)}(t) \hat b +f_5^{(n)}(t)\hat b^{\dagger}+f_6^{(n)}(t) 
\end{aligned}
\end{equation}
the phonon number operator in Heisenberg's representation is then given by
\begin{equation}\label{eq:NHeis}
\begin{aligned}
& \hat N(n,t) = \left(f_1^{(n)} f_4^{(n)}+f_2^{(n)} f_5^{(n)}\right)\hat N +\left( f_2^{(n)} f_5^{(n)}+f_3^{(n)}f_6^{(n)}\right){\cal I}+ f_1^{(n)} f_5^{(n)} \hat b^{\dagger 2} \\ & + f_2^{(n)} f_4^{(n)}\hat b^2  + \left(f_1^{(n)} f_6^{(n)} +f_3^{(n)}f_5^{(n)} \right)\hat b^{\dagger}+ \left(f_2^{(n)} f_6^{(n)} + f_3^{(n)} f_4^{(n)} \right)\hat b
\end{aligned}
\end{equation}
where we have omitted the temporal dependence of the functions $f_k^{(n)}$.
When we take the average between mechanical coherent states, $|\Gamma\rangle$ we obtain,
\begin{equation}\label{eq:Navercoherente}
\begin{aligned}
& \langle \hat{N}(t)\rangle_{\Gamma} = \left(f_1^{(n)} f_4^{(n)}+f_2^{(n)} f_5^{(n)}\right)|\Gamma|^2 +\left( f_2^{(n)} f_5^{(n)}+f_3^{(n)}f_6^{(n)}\right)+ f_1^{(n)} f_5^{(n)} \Gamma^{*2}  \\ & + f_2^{(n)} f_4^{(n)}\Gamma^2  + \left(f_1^{(n)} f_6^{(n)} +f_3^{(n)}f_5^{(n)} \right)\Gamma^{*}+ \left(f_2^{(n)} f_6^{(n)} + f_3^{(n)} f_4^{(n)} \right)\Gamma \\ &  \equiv \Psi_0^{(n)} |\Gamma|^2 +\Psi_1^{(n)} +\Psi_2^{(n)} \Gamma^{* 2}+\Psi_3^{(n)} \Gamma^2+\Psi_4^{(n)} \Gamma^{*} +\Psi_5^{(n)} \Gamma
\end{aligned}
\end{equation}

\begin{figure}[hbtp]
    \centering
    \includegraphics[width = \linewidth]{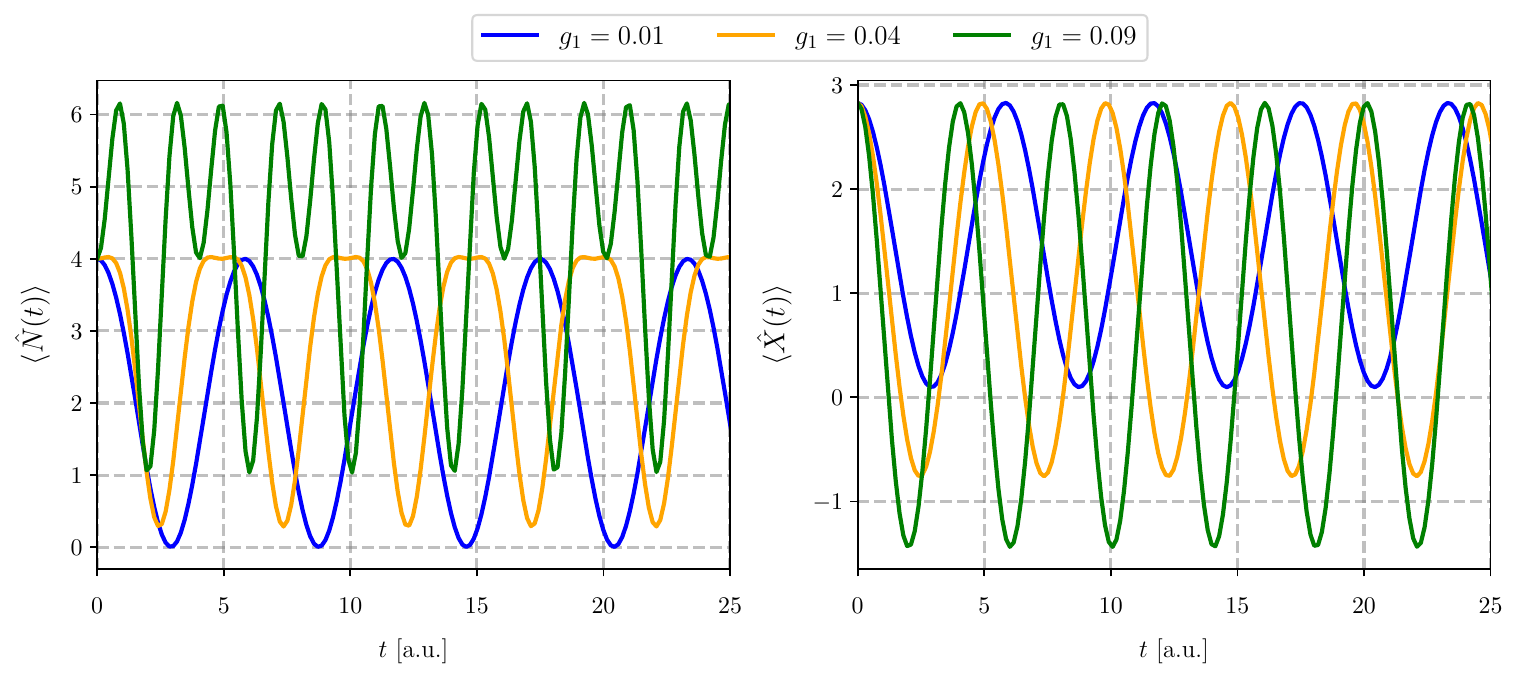}
    \caption{The average value of the phonon number operator (left) and the position quadrature (right) for an initial state of the field with $n=4$, and a coherent state of the mechanical oscillator with $\Gamma=2$. The linear coupling constant is $g_0=0.3$ and the quadratic coupling constants are: $g_1=0.01$ (blue), $g_1=0.04$ (orange) and $g_1=0.09$ (green). The oscillator's frequency is $\omega_m = 1$.}
    \label{Naverlincuad}
\end{figure}

Using the transformations given by Eqs.\ref{eq:transbd} and \ref{eq:transb} we can obtain $\hat{N}^{2}(t)$ and its average. With these, we can evaluate the Mandel parameter
\begin{equation}\label{eq:Mandel}
Q_{\Gamma} = \frac{\langle \hat{N}^2(t)\rangle_{\Gamma} -\langle \hat{N}(t)\rangle_{\Gamma}^2}{\langle \hat{N}(t)\rangle_{\Gamma}}
\end{equation}
the numerator is:
\begin{equation}
\begin{aligned}
\langle \hat{N}^2(t)\rangle_{\Gamma} -\langle \hat{N}(t)\rangle_{\Gamma}^2 & =  |\Gamma|^2(\Psi_0^{(n)2}+4\Psi_2^{(n)} \Psi_3^{(n)})+ 2\Gamma^{*2}\Psi_0^{(n)} \Psi_3^{(n)} + \Gamma^{*}(\Psi_0^{(n)}\Psi_4^{(n)}+2\Psi_5^{(n)}\Psi_2^{(n)}) \\ & +  2\Gamma^2\Psi_2^{(n)} \Psi_0^{(n)} + \Gamma(2\Psi_3^{(n)} \Psi_4^{(n)}+\Psi_5^{(n)} \Psi_0^{(n)})+ \Psi_3^{(n)} \Psi_2^{(n)}+\Psi_5^{(n)} \Psi_4^{(n)}
\end{aligned}
\end{equation}
and the denominator is given in Eq.~\ref{eq:Navercoherente}. \\
The quadratures 
$\hat{X} =\frac{1}{\sqrt{2}}\left(\hat{b}+\hat{b}^{\dagger} \right), \ \hat{P} = \frac{\mathrm{i}}{\sqrt{2}}\left( \hat{b}^{\dagger} -\hat{b} \right)$ in the Heisenberg representation are given by:
\begin{equation}
\begin{aligned}
 \hat{X}(n,t) &= \frac{1}{\sqrt{2}}\left((f_2^{(n)} + f_4^{(n)})\hat{b} + (f_1^{(n)}+f_5^{(n)})\hat{b}^{\dagger} + f_3^{(n)}f_6^{(n)} \right)\\
\hat{P}(n,t) &= \frac{\mathrm{i}}{\sqrt{2}}\left((f_4^{(n)}-f_2^{(n)})\hat{b}^{\dagger} +(f_5^{(n)} -f_1^{(n)})\hat{b} +f_6^{(n)} -f_3^{(n)} \right)
\end{aligned}
\end{equation}
and their dispersions evaluated between number field states $|n\rangle$ and mechanical coherent states $|\Gamma\rangle$  are: 
\begin{equation}
\Delta X(n,t) = \sqrt{\frac{1}{2}(f_1^{(n)}+f_5^{(n)})(f_2^{(n)}+f_4^{(n)})}, \ \Delta P(n,t) = \sqrt{-\frac{1}{2}(f_4^{(n)}-f_2^{(n)})(f_5^{(n)}-f_1^{(n)})} \ 
\end{equation}
When we consider a coherent state $|\alpha\rangle$ for the cavity field we obtain:
\begin{equation}
    \Delta X(\alpha,t) = e^{-|\alpha|^2}\sum_{k=0}^{\infty} \frac{|\alpha|^{2k}}{k!} \Delta X(k,t), \ \ \ \Delta P(\alpha,t) = e^{-|\alpha|^2}\sum_{k=0}^{\infty} \frac{|\alpha|^{2k}}{k!} \Delta P(k,t)
\end{equation}
In figure \ref{Naverlincuad} we show, in the left panel, the average phonon number operator $\bar{N}= \langle \hat{N}(n,t)\rangle$ taken between a coherent mechanical oscillator state with $\Gamma=2$ and a number state for the field with $n=4$ for a fixed value of the linear coupling $g_0=0.3$, that corresponds to a cooling of the mechanical oscillator when the quadratic term is absent (see Fig.~\ref{Naverage}), and several values of the quadratic coupling $g_1$.    
The cavity frequency is $\omega_c=10\omega_m$ and the effective frequency of the mechanical oscillator is different in each case. For $g_1=0.01$ the effective frequency is $\omega_m^{(eff)}= 1.08 \omega_m$, for $g_1=0.04$ it is, $\omega_m^{(eff)}=1.32 \omega_m$ and for $g_1=0.09$ it is $\omega_m^{(eff)} = 1.54\omega_m$.  \\
The average phonon number is a periodic function of time with a period $T^{(eff)}=2\pi/\omega_m^{(eff)}$. When the quadratic coupling is small (blue line) the average phonon number is a decreasing periodic function of time going from $\bar{N} =4$ to $\bar{N}=0$ a result similar to that found in figure \ref{Naverage} though the minimum is now closer to zero.  Increasing the quadratic coupling to $g_1=0.04$, we again see a periodic decreasing function with a different period and a kind of plateau at $\bar{N}=4$ (orange line). Notice that this plateau is absent in figure \ref{Naverage}. Increasing even more the quadratic coupling to $g_1=0.09$, we notice a radical change. The average number of phonons starts as an increasing function, going up to $\bar N=6$ then, it decreases to $\bar N=1$ and increases again up to 6; thereafter, it oscillates with smaller amplitude going from $\bar N=6$  to $\bar N= 4$ and then up to 6 once more. This conduct is repeated periodically. It is clear that the evolution of the average phonon number with linear and quadratic couplings is much more subtle than that found for the linear or the quadratic cases. The effect of keeping linear and quadratic couplings is not the sum of the independent contributions.  In the right panel of figure \ref{Naverlincuad} we show the average value of the position quadrature for $g_0=0.3$ and several values for the quadratic coupling $g_1$. Recall that for this value of the linear coupling and without the quadratic one, the evolution of the position quadrature was an oscillatory function evolving from $X\simeq 3$ to $X\simeq 0.7$ with a period of $2\pi/\omega_m$. Here we see that with $g_1=0.01$ the position, quadrature has a larger amplitude of oscillation and a shorter period, now around. Increasing even more, the quadratic coupling increases the amplitude of the oscillations and shortens their period.

\begin{figure}[hbtp]
    \centering
    \includegraphics[width = \linewidth]{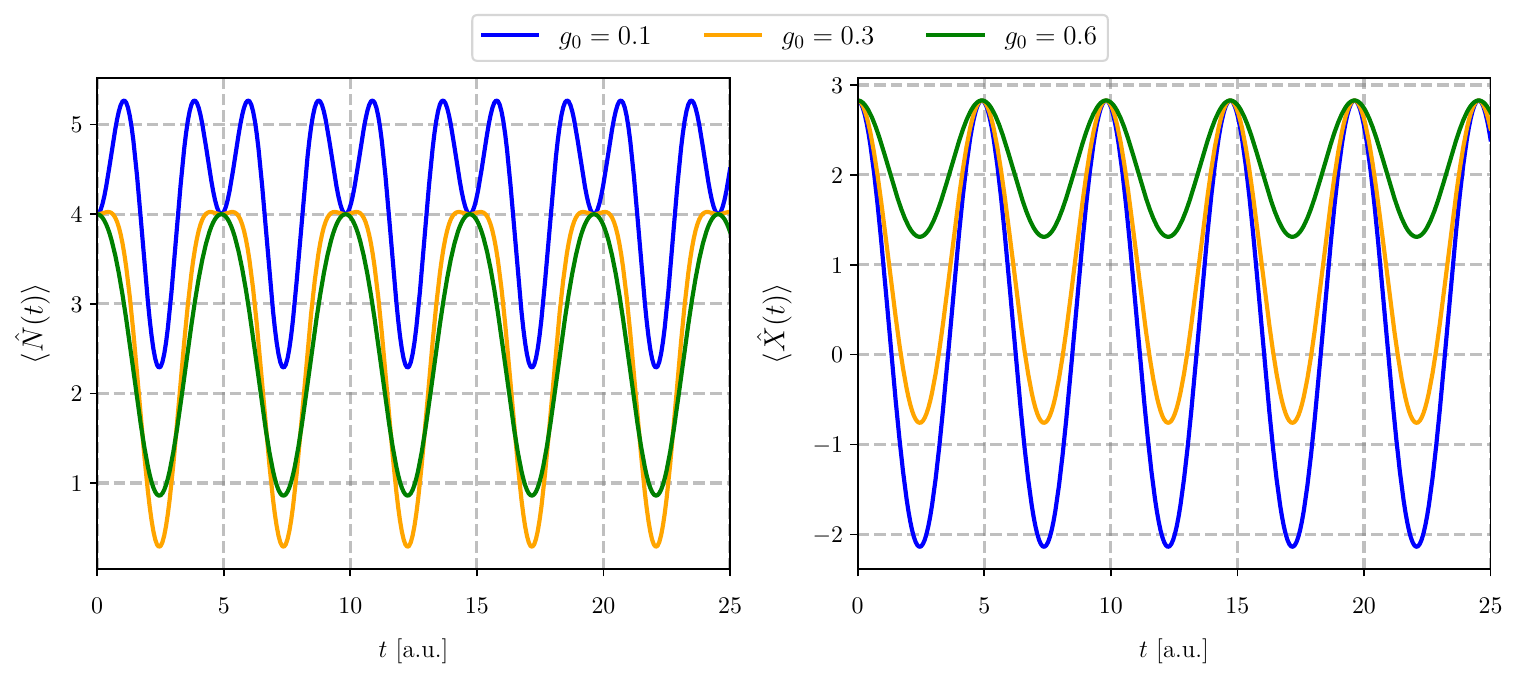}
    \caption{The average value of the phonon number operator (left) and the position quadrature (right) for an initial state with $n=4$, $\Gamma = 2$, a fixed quadratic coupling $g_1=0.04$ and linear couplings:  $g_0=0.1$ (blue), $g_{0} = 0.3$, (orange) and $g_0=0.6$ (green). The oscillator's frequency is $\omega_m = 1$.}
    \label{Naverlincuadb}
\end{figure}

In figure \ref{Naverlincuadb} we show the average phonon number operator and the average position quadrature taken between a coherent mechanical oscillator state $\Gamma=2$ and a number state for the field with $n=4$ with a fixed value for the quadratic coupling $g_1=0.04$ and several values of the linear coupling $g_0$.
In this case, the plots in the figure have the same frequency, $\omega= 1.32 \omega_m$ because the quadratic coupling is fixed.  When we had only a linear coupling, the cases with $g_0=0.1$ and $g_0=0.3$ corresponded to cooling of the oscillator; that is, a decrement in the phonon number operator average. Here we see that, due to the nonlinear coupling, for $g_0=0.1$ (blue line) the oscillator starts at $\bar{N}=4$ and initially increases its value up to $\bar{N}\approx 5.2$, then decreases it to $\bar{N}\approx 2.4$   and increases again up to $\bar{N}\approx 5.2$. This conduct is similar to that shown in figure \ref{Naverlincuad} for the case with $g_1=0.09$ and $g_0=0.3$. For $g_0=0.3$ (orange line) $\bar{N}$ is a decreasing periodic function starting at $\bar{N}=4$ and decreasing to $\bar{N}\approx 0.3$. 
Finally, as $g_0=0.6$ in the linear case we found a heating behavior, but here we see a cooling behavior with $1\leq \bar{N}\leq 4$ a conduct opposed to what we had found in the linear case. 

If we take the average of Eq.~{\ref{eq:NHeis}} between  a field coherent state $|\alpha\rangle$ and a mechanical oscillator coherent state $|\Gamma\rangle$, we obtain:
\begin{equation}
    \langle \alpha,\Gamma|\hat{N}(n,t)|\alpha,\Gamma\rangle = e^{-|\alpha|^2} \sum_{k=0}^{\infty} \frac{|\alpha|^{2k}}{k!} \langle \hat{N}(k,t)\rangle  
\end{equation}

\begin{figure}[htbp]
    \centering
    \includegraphics[width = \linewidth]{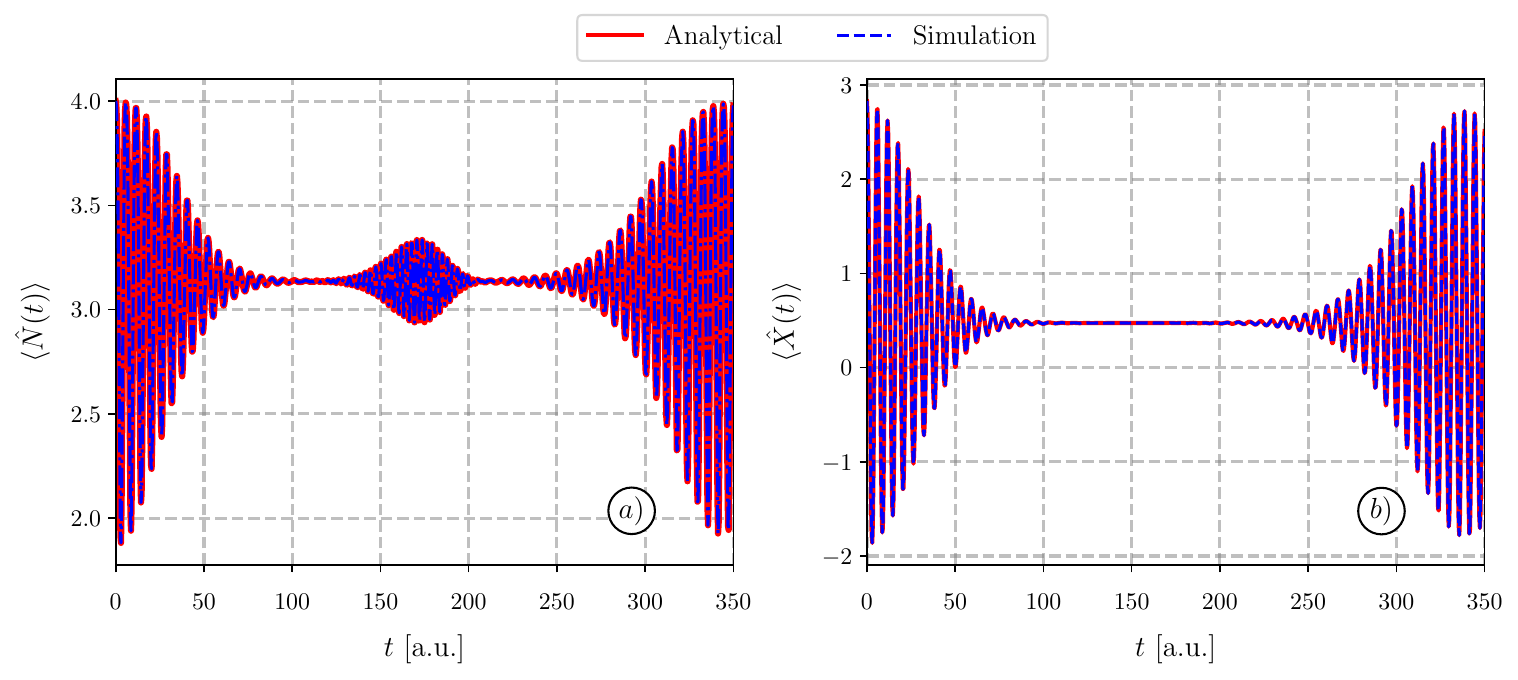}
    \caption{Comparison between analytical method and numerical simulation with linear and quadratic couplings. Panel {\em a)} shows the mean expectation value of the phonon number operator and panel {\em b)} the average value of the position quadrature, for an initial state $\ket{\psi(0)} = \ket{\alpha,\Gamma} = \ket{2,2}$. The Hamiltonian parameters are: $\omega_{c} = 10.0$, $\omega_{m} = 0.1\omega_{c}$, $g_{0} = 0.1$, $g_{1}=g_{0}^{2}$.}
    \label{fig:Nq_nonfor}
\end{figure}

In figure \ref{fig:Nq_nonfor} we show the average phonon number operator and the position quadrature for the case with $g_0=0.1$ and $g_1=0.01$. Figures correspond to the initial state $|\Psi(0)\rangle = |\alpha,\Gamma\rangle = |2,2\rangle$.  We show the results of our analytical method in full line (red) and those obtained by a numerical simulation in broken line (blue). Since in this case, the analytical method is exact, both calculations should yield the same result for a converged simulation. We did this comparison to test our numerical simulation. 
Comparison with figure \ref{Navercoh} shows that the principal effect of the linear coupling is a displacement in $\bar{N}(t)$. When there is no linear coupling, $\bar{N}$ starts at 4 and {\em increases} up to $\approx 4.6$, after some time, we see the appearance of a first collapse at $\bar{N}\approx 4.3$ and subsequent revivals and collapses. The length of the collapses increasing and their amplitude decreasing with time. When the linear coupling is different from zero, $\bar{N}$ starts at 4 and {\em diminishes} to around 2. The first collapse appears at about the same time as in the case with no linear coupling; however, the first revival is much less intense than that of figure \ref{Navercoh}. After this, there is a second collapse and a revival whose intensity is similar to that shown in figure \ref{Navercoh}. Then, there is a sequence of collapses and revivals (one small, the next larger).  In panel {\em b)} we see that the evolution of the quadrature shows a similar conduct as in the case with no linear term; however, the collapse is now around 0.8 and when there is no linear term it is around zero. 

\begin{figure}[hbtp]
    \centering
    \includegraphics[width = \linewidth]{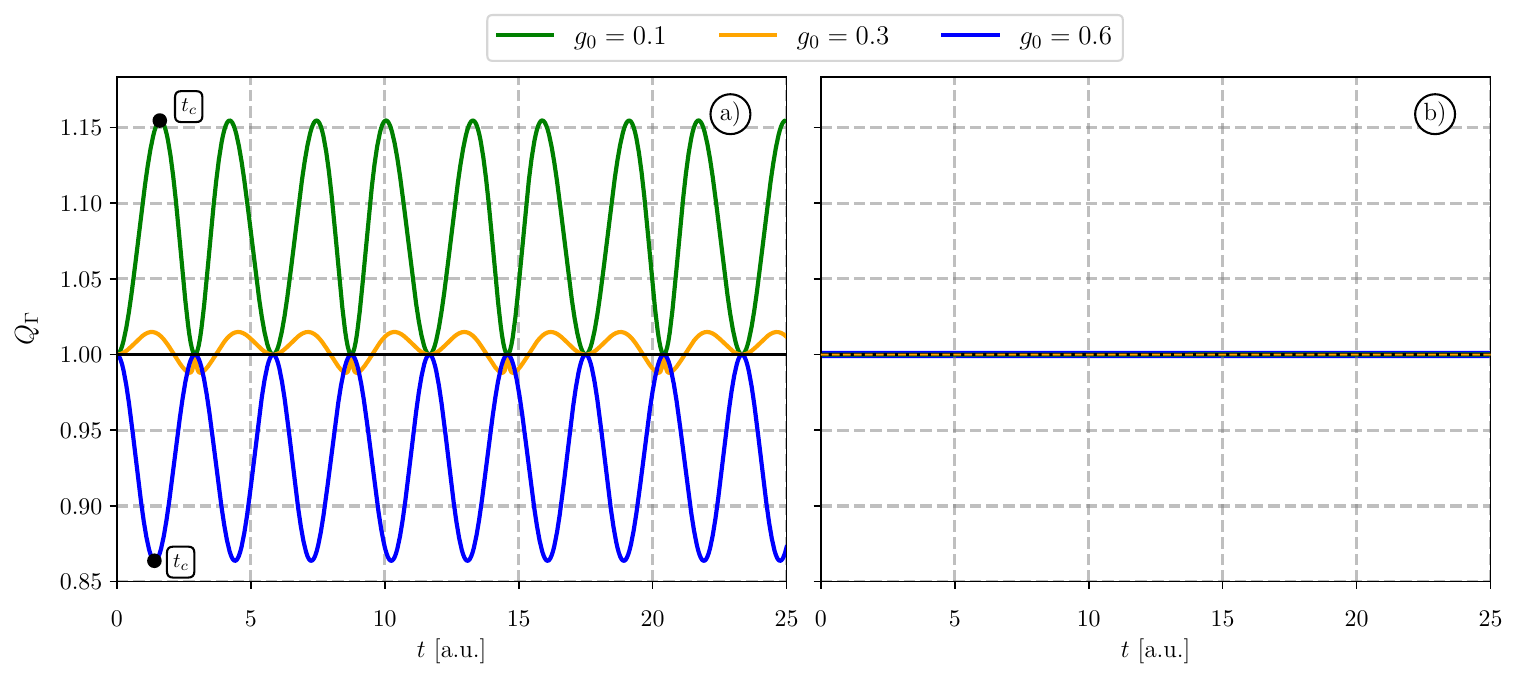}    \caption{Temporal evolution of the Mechanical Mandel parameter $Q_{\Gamma}$ for a system with $n=4$ photons and a mechanical coherent state with $\Gamma=2$. Left figure with nonlinear coupling $g_1=0.01$ and $g_0=0.1$ (green), $g_0=0.3$ (orange) and $g_0=0.6$ (blue). Right figure with $g_1=0$ and the same values for $g_0$. The oscillator's frequency is $\omega_m = 1$.}
    \label{Mandel1}
\end{figure}

In figure \ref{Mandel1} we show the temporal evolution of the Mandel $Q$ parameter of the mechanical oscillator. In the left panel, for a fixed value of the quadratic coupling $g_1=0.01$ and the same values for the linear coupling as those used in figure \ref{Naverage}. We can see there that for $g_0=0.1$ and $g_0 = 0.3$ the average number of phonons is a decreasing periodic function of time, recall that, in figure \ref{Naverage} there is no quadratic coupling. When we include a small value for the quadratic coupling, for instance $g_1=0.01$,  the average value of the phonon number operator remains practically the same. We see in figure \ref{Mandel1} that for these values of the linear coupling, the Mandel parameter is larger than one, corresponding to a super-Poissonian distribution. When the linear coupling is increased to $g_0=0.6$, the average number of phonons becomes an increasing periodic function of time and the corresponding Mandel parameter is always smaller than one, as corresponds to a sub-Poissonian statistics. Then, there is a change in the statistical behavior of the system as a function of the intensity of the linear coupling. It goes from cooling to heating and from super Poissonian to sub Poissonian statistics. In the right panel, we show the case without quadratic coupling, and the same values for the linear coupling parameter $g_0$. We see that whenever the quadratic coupling is zero, the Mandel parameter is equal to one as corresponds to a coherent state, with no regard to the value for the linear coupling.
\begin{figure}[H]
    \centering
    \includegraphics[width = \linewidth]{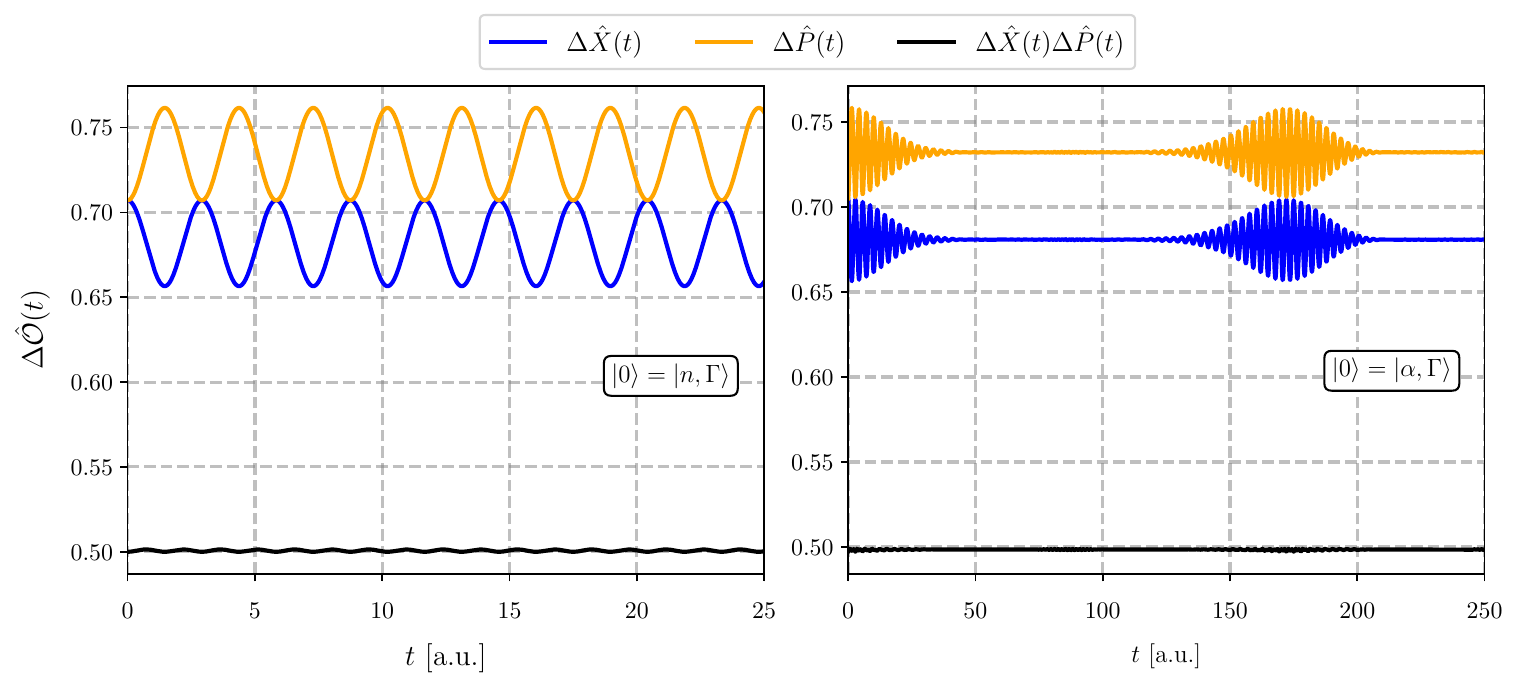}
    \caption{Temporal evolution of the dispersions in the mechanical quadratures $\Delta \hat{X}(n,t)$ (blue), $\Delta \hat{P}(n,t)$ (orange) and their product (black) for a system with a number state for the field  $n=4$ (left) and $\Delta \hat{X}(\alpha,t)$ (blue), $\Delta \hat{P}(\alpha,t)$ (orange) for a coherent  state for the field with $\alpha = 2$ (right). For the mechanical oscillator, we used a coherent state with $\Gamma=2$. The Hamiltonian parameters are $g_0=0.1$, $g_1=0.01$, $\omega_c=10$, $\omega_m=1$.}
    \label{Dispersiones}
\end{figure}

In \cite{jahne09} the authors studied the squeezed states of a mechanical oscillator in an optomechanical system driven by a coherent field and a squeezed vacuum field. Here, we present, in the left panel of figure \ref{Dispersiones}, the temporal evolution of the dispersions of the quadratures of the mechanical oscillator $\Delta X(n,t)$ (blue) and $\Delta P(n,t)$ (orange). For these calculations we used $g_0=0.1$, $g_1=0.01$,   $n=2$, $\Gamma=2$.  We see that due to the non-zero value of the quadratic coupling, there is squeezing present. The quadrature $\hat{X}(n,t)$ takes values below those allowed by Heisenberg's principle, while the quadrature $\hat{P}(n,t)$ takes values above. Their product is a constant with the minimum allowed value as  corresponds to a coherent state. We see also that the dispersions in both quadratures are equal at multiples of $ T_m/2$, with $T_m$ the period of the mechanical oscillator. When we make $g_1=0$ we obtain the same constant value for each quadrature and their product is the minimum value allowed by Heisenberg's principle, the result in agreement with that obtained employing the Mandel parameter.
In the right panel of the figure we see the evolution of the quadratures $\Delta \hat{X}(\alpha,t)$ and $\Delta \hat{P}(\alpha,t)$ for an initial coherent state of the field $|\alpha\rangle$.  We see that due to the interferences of the number states in the distribution of the coherent field state, the squeezing now presents collapses and revivals. The dispersion in the position and momentum quadratures retain the conduct we found for a number field state, that is, the dispersion in the position quadrature is smaller than the allowed by Heisenberg's principle while that of the momentum is larger their product a constant value corresponding to a coherent state.

\section{Forced optomechanical system with linear and quadratic couplings}
The Hamiltonian describing the forced optomechanical system with linear and quadratic couplings is:
\begin{equation}\label{eq:Hfull}
\begin{aligned}
\frac{\hat{H}}{\hbar} & =  (\omega_c+g_1\omega_m)\hat{n} +\omega_m\hat{N} +2g_1 \omega_m \hat{n}\hat{N} -g_0\omega_m\hat{n}(\hat{b}+\hat{b}^{\dagger}) \\ &  + g_1 \omega_m \hat{n}(\hat{b}^{\dagger 2}+\hat{b}^2) + \Omega \cos(\omega_d t)(\hat a +\hat a^\dagger). 
\end{aligned}
\end{equation}
In this case, the photon number operator is not a constant of motion. 
We choose as unperturbed Hamiltonian
\begin{equation}
    \hat{H}_0 = \hbar (\omega_c+g_1\omega_m)\hat{n} + \hbar \omega_m \hat{N} \equiv \hbar \omega_c' \hat{n}+\hbar \omega_m \hat{N}
\end{equation}
the interaction picture Hamiltonian is:
\begin{equation}
\begin{aligned}
 \frac{\hat{H}_I(t)}{\hbar} & =  2g_1\omega_m \hat{n}\hat{N} -g_0\omega_m\hat{n}(\hat{b}e^{-i\omega_m t}+\hat{b}^{\dagger}e^{\mathrm{i}\omega_m t}) \\ & +  g_1\omega_m \hat{n}(\hat{b}^{\dagger 2}e^{2\mathrm{i}\omega_m t}+\hat{b}^2 e^{-2\mathrm{i}\omega_m t}) + \Omega \cos(\omega_d t)(\hat{a} e^{-\mathrm{i}\omega_c' t}+\hat{a}^{\dagger}e^{\mathrm{i}\omega_c' t})
\end{aligned}
\end{equation}
now we write $\hat{H}_I(t) = \hat{H}_I^{(1)}(t)+\hat{H}_I^{(2)}(t)$ with:
\begin{equation}
    \hat{H}_I^{(1)}(t) = \hbar \Omega \cos(\omega_d t)(\hat{a}e^{-i\omega_c' t}+\hat{a}^{\dagger} e^{i\omega_c' t})
\end{equation}
the corresponding time evolution operator is
\begin{equation}\label{eq:Uint}
    \hat{U}_I^{(1)}(t) = e^{\alpha_1(t) \hat{a}^{\dagger}} e^{\alpha_2(t) \hat{a}} e^{\alpha_3(t)}
\end{equation}
with the functions $\alpha_i(t)$ such that:
\begin{equation}
\begin{aligned}
    \dot \alpha_1(t) & =  -\mathrm{i} \Omega e^{\mathrm{i}\omega_c' t} \cos{\omega_d t} \\
    \dot \alpha_2(t) & =  -\mathrm{i} \Omega e^{-\mathrm{i} \omega_c' t} \cos{\omega_d t} \\
    \dot \alpha_3(t) &= \alpha_1(t) \dot \alpha_2(t)
\end{aligned}
\end{equation}
these equations can be integrated, and we obtain
\begin{equation}\label{eq:alpha1}
    \alpha_1(t) = \frac{\Omega}{\omega_c'^2-\omega_d^2}\left(\omega_c'+e^{\mathrm{i}\omega_c' t}(-\omega_c' \cos{\omega_d t}+\mathrm{i}\omega_d \sin{\omega_d t}) \right)
\end{equation}
\begin{equation}
\alpha_2(t) = \frac{\Omega}{\omega_c'^2-\omega_d^2}\left(-\omega_c'+e^{-\mathrm{i}\omega_c' t}(\omega_c' \cos{\omega_d t}+\mathrm{i}\omega_d \sin{\omega_d t}) \right)
\end{equation}
and we see that $\alpha_1(t) = -\alpha_2^{*}(t)$ so that the time evolution operator given in Eq.~\ref{eq:Uint} can be written in terms of a displacement operator.\\
Now we have to transform the rest of the interaction
\begin{equation}
    \tilde{H}_I^{(2)}(t) = \hat{U}_I^{(1)\dagger}(t)\hat{H}_I^{(2)}(t)\hat{U}_I^{(1)}(t)
\end{equation}

Using the explicit form of $\hat U_I^{(1)}(t)$ we obtain
\begin{gather}
\hat{a}^{\dagger}(t) =  \hat U_I^{(1)\dagger}(t) \hat{a}^{\dagger}\hat{U}_I^{(1)}(t)  =  \hat a^{\dagger}-\alpha_2(t) \\
\hat{a}(t) = \hat U_I^{(1)\dagger}(t) \hat{a}\hat{U}_I^{(1)}(t)  =  \hat a + \alpha_1(t)
\end{gather}
and the photon number operator in the Heisenberg representation is:
\begin{equation}\label{eq:nheis}
    \hat n(t) = \hat n +\alpha_1(t) \hat a^{\dagger} +\alpha_1^{*}(t) \hat a + |\alpha_1(t)|^2
\end{equation}
   The transformed interaction takes the form:
   \begin{equation}\label{eq:hint}
\frac{\tilde{H}_I^{(2)}(t)}{\hbar} = \omega_m \hat{n}(t)\left(2g_1\hat{N} -g_0(\hat{b}e^{-\mathrm{i}\omega_m t}+\hat{b}^{\dagger}e^{\mathrm{i}\omega_m t})+g_1(\hat{b}^{2}e^{-2\mathrm{i}\omega_m t}+\hat{b}^{\dagger 2}e^{2\mathrm{i}\omega_m t}) \right)
   \end{equation}
and we see that the transformed Hamiltonian contains terms like
\begin{equation}
    \hat{n}(t)\left(\hat{b}e^{-\mathrm{i}\omega_m t}+\hat{b}^{\dagger}e^{\mathrm{i}\omega_m t} \right) = \left( \hat{n} +|\alpha_1(t)|^2\right)(\hat{b}e^{-\mathrm{i}\omega_m t}+\hat{b}^{\dagger} e^{\mathrm{i}\omega_m t}) +\left(\alpha_1(t)\hat{a}+\alpha_1^{*}(t)\hat{a}^{\dagger}\right)
(\hat{b}e^{-\mathrm{i}\omega_m t}+\hat{b}^{\dagger} e^{\mathrm{i}\omega_m t}).
\end{equation}
The first term is the one we had for the field-mechanical oscillator coupling in the linear case, plus a time-dependent displacement $|\alpha_1(t)|^2$. We obtained its solution in section \ref{sec:linearcoupling}. The second term involves the exchange of quanta between the field and the mechanical oscillator, with a coupling  given in Eq.~{\ref{eq:alpha1}}. \\
Considering now the quadratic term, we have:
\begin{equation}
\begin{aligned}
    \hat{n}(t)\left(\hat{b}^{2}e^{-2\mathrm{i}\omega_m t}+\hat{b}^{\dagger 2}e^{2i\omega_m t}\right) &=  \left( \hat{n} +|\alpha_1(t)|^2\right)\left(\hat{b}^{2}e^{-2\mathrm{i}\omega_m t}+\hat{b}^{\dagger 2}e^{2\mathrm{i}\omega_m t}\right)\\ 
    &+ \left(\alpha_1(t)\hat{a}+\alpha_1^{*}(t)\hat{a}^{\dagger}\right)\left(\hat{b}^{2}e^{-2\mathrm{i}\omega_m t}+\hat{b}^{\dagger 2}e^{2\mathrm{i}\omega_m t}\right)
\end{aligned}
\end{equation}
the first term is the one we got for the quadratic coupling between the field and the mechanical oscillator plus a time-dependent displacement.  Its solution was obtained in section \ref{sec:quadratic}. The second term corresponds to an exchange between a photon and a pair of phonons and the coupling is similar to the one we found for the linear case. Finally, we have,
\begin{equation}
    \hat{n}(t)\hat{N} =  \left( \hat{n} +|\alpha_1(t)|^2\right)\hat{N} +  \left(\alpha_1(t)\hat{a}+\alpha_1^{*}(t)\hat{a}^{\dagger}\right) \hat{N}
\end{equation}
the first term is similar to the one we had in section \ref{sec:linplusquad} and the second term involves a coupling between the number of excitations in the mechanical oscillator and the field operators $\hat{a}$, $\hat{a}^{\dagger}$ and the intensity of the coupling is also given by Eq.~\ref{eq:alpha1}. 

Several works have been published emphasizing the strong coupling regime of a single photon \cite{jie14, rabl11,nunnenkamp11}. To consider this particular situation, we are going to keep only a few photons in the cavity; then, we have to consider the case where the forcing term satisfies conditions $\omega_d \ll \omega_c^{'} $ and $\Omega < \omega_m$. Under these, we approximate the photon number operator by
\begin{equation}\label{naver}
    \hat{n}(t) \approx \hat{n} +|\alpha_1(t)|^2.
\end{equation}
\begin{figure}[H]
    \centering
    \includegraphics[width = \linewidth]{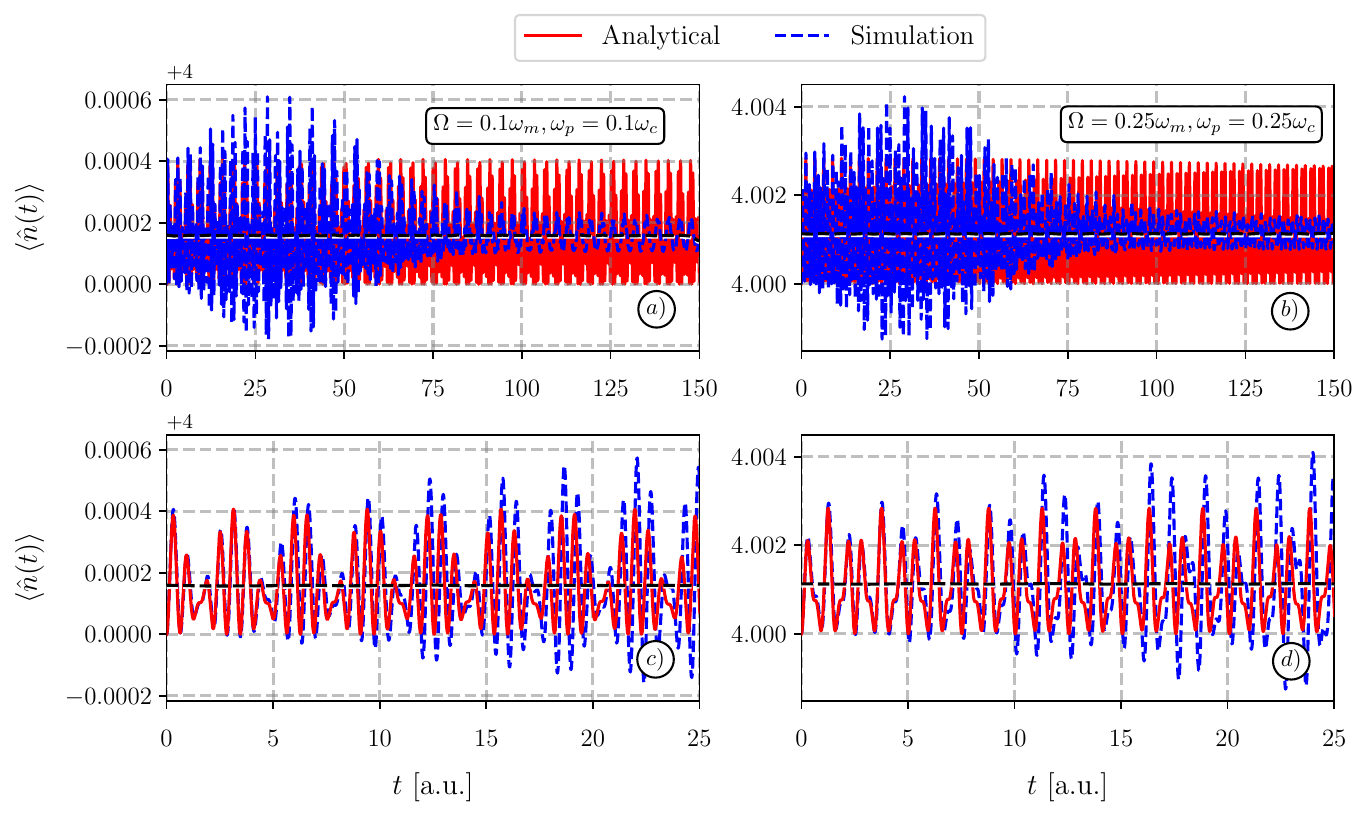}
    \caption{Comparison between the approximation for the forced number of photons given by Eq.~\eqref{naver} and a numerical simulation with Hamiltonian parameters: $\omega_{c} = 10$, $\omega_{m} = 0.1\omega_{c}$, $g_{0} = 0.1$, $g_{1} = g_{0}^{2}$. Following the conditions $(\omega_{d}, \Omega) \ll \omega_{c}$, we use $\Omega = 0.25\omega_{m}$ and $\omega_{d} = 0.25\omega_{c}$. We see a good agreement between the analytical approximation and the simulation at short times when we consider the moving average, black and white lines, of both lines.}
    \label{naverf_comp}
\end{figure}
In figure \ref{naverf_comp} we show the temporal evolution of the average photon number operator given by Eq.~\ref{naver} and a numerical simulation using the full Hamiltonian given by Eq.~\ref{eq:Hfull} with Hamiltonian parameters $\omega_c=10$, $\omega_m=0.1\omega_c$, $g_0=0.1$, $g_1=g_0^2$. On the left-hand side $\omega_d=0.1\omega_c$ and $\Omega=0.1\omega_m$. In the panel, $a)$ we plot a time interval of about 25 periods of the mechanical oscillator. We notice that the analytical result is a periodic function of time going from $\langle \hat{n}(t)\rangle = 4$ to a maximum value of $\langle \hat{n}(t)\rangle = 4.0006$, that is, the change in the average photon number is only of the order of $10^{-4}$ for this set of parameters; the numerical result shows a different structure (collapses and revivals), however, a temporal average of the analytic and the numerical results would yield to similar results. In the panel $b)$ we plot the same time interval but with $\Omega= 0.25 \omega_m$ and $\omega_p =0.25 \omega_c$ in this case, the change in the average photon number is of the order  of $10^{-3}$. In the second row,  we show the first few periods of the evolution, we can see that the approximate expression shows a good agreement with the numerical simulation.

 Since with this approximation, the photon number operator commutes with the Hamiltonian, we can replace it by its average value. 
 In so doing, the only operators in Eq.\ref{eq:hint} are those corresponding to the mechanical oscillator $\{\hat b, \hat b^{\dagger}, \hat N, \hat b^2, \hat b^{\dagger 2}, {\cal I} \}$. These form a finite Lie algebra, and we can write the time evolution operator corresponding to the Hamiltonian in a product form, as in Eq.~\ref{eq:Unl}.
\begin{equation}
\hat{U}_I^{(2)}(t) = \prod_{k=1}^6 e^{\beta_k^{(n)}(t) \hat X_k}.
\end{equation}
Within this approximation, the full-time evolution operator is:
 \begin{equation}
     \hat{U}(t) = \hat{U}_0(t) \hat{U}_I^{(1)}(t) \hat{U}_I^{(2)}(t)
 \end{equation}
The phonon creation-annihilation operators in the Heisenberg representation are:
\begin{gather}
    \hat{b}(t) = \hat{U}^{\dagger}(t)\hat{b}\hat{U}(t)= t_1\hat{b}+t_2\hat{b}^{\dagger}+t_3\\
    \hat{b}^{\dagger}(t) = \hat{U}^{\dagger}(t)\hat{b}^{\dagger}\hat{U}(t) =t_4 \hat{b}^{\dagger} + t_5 \hat{b} + t_6
\end{gather}
where
\begin{gather}
        t_1 = e^{-i\omega_m t}\left(e^{\beta_3} - 4\beta_1\beta_5 e^{-\beta_3}\right), \ t_2 = e^{-\mathrm{i}\omega_m t} 2\beta_1 e^{-\beta_3}, \ t_3 = e^{-\mathrm{i}\omega_m t}\left(\beta_2-2\beta_1\beta_4e^{-\beta_3}\right)\\
        t_4 = e^{\mathrm{i}\omega_m t}e^{-\beta_3}, \ t_5 = -2\beta_5 e^{\mathrm{i}\omega_m t} e^{-\beta_3}, \ t_6 = -\beta_4 e^{\mathrm{i}\omega_m t}e^{-\beta_3}
\end{gather}
The phonon number operator is then,
\begin{equation}
\begin{aligned}
    \hat{N}(t) & =  \left(|t_1|^2+|t_2|^2\right)\hat{N} +t_1^{*} t_2 \hat{b}^{\dagger 2}+\left(t_1^{*} t_3 + t_2 t_3^{*}\right)\hat{b}^{\dagger} \\ & +\left(t_2^{*} t_3 + t_3^{*}t_1\right)\hat{b} + t_1t_2^{*} \hat{b}^2 + |t_2|^2+|t_3|^2
\end{aligned}
\end{equation}
and the quadratures 
\begin{equation}
\begin{aligned}
    \hat{X}(t) & = \frac{1}{\sqrt{2}}\left(\hat{b}(t) +\hat{b}^{\dagger}(t) \right) = \frac{1}{\sqrt{2}}\left((t_1+t_2^{*})\hat{b} +(t_2+t_1^{*})\hat{b}^{\dagger}+ t_3 + t_3^{*} \right) \\
    &  \equiv \frac{1}{\sqrt{2}}\left(\psi_1 \hat{b} +\psi_2 \hat{b}^{\dagger} +\psi_3 \right)
\end{aligned}
\end{equation}
and
\begin{equation}
\begin{aligned}
    \hat{P}(t) & = \frac{\mathrm{i}}{\sqrt{2}}\left(\hat{b}^{\dagger}(t) -\hat{b}(t) \right) =  \frac{\mathrm{i}}{\sqrt{2}}\left((t_1^{*}-t_2) \hat{b}^{\dagger} +(t_2^{*}-t_1) \hat{b} +t_3^{*}-t_3 \right)  \\
    &  \equiv \frac{\mathrm{i}}{\sqrt{2}}\left(\psi_4 \hat{b}^{\dagger} +\psi_5 \hat{b} +\psi_6 \right)
\end{aligned}
\end{equation}
whose average values between mechanical coherent states $|\Gamma\rangle$ are easy to calculate
\begin{equation}
    \langle \Gamma|\hat{X}(t)|\Gamma\rangle = \frac{1}{\sqrt{2}}\left(\psi_1 \Gamma +\psi_2 \Gamma^{*}+\psi_3 \right), \ \langle \Gamma|\hat{P}(t)|\Gamma\rangle = \frac{\mathrm{i}}{\sqrt{2}}\left(\psi_4 \Gamma^{*}+\psi_5 \Gamma +\psi_6 \right)
\end{equation}
and after some algebra, we can also evaluate their dispersions
\begin{equation}
    \begin{aligned}
        \Delta X(t) &= \sqrt{\langle \Gamma|\hat{X}^2(t)|\Gamma\rangle -\langle \Gamma|\hat{X}(t)|\Gamma\rangle^2} =\sqrt{\frac{1}{2}\psi_1 \psi_2},\\
        \Delta P(t) &= \sqrt{\langle \Gamma|\hat{P}^2(t)|\Gamma\rangle -\langle \Gamma|\hat{P}(t)|\Gamma\rangle^2} = \sqrt{-\frac{1}{2}\psi_4\psi_5}
    \end{aligned}
\end{equation}
\begin{figure}[H]
    \centering
    \includegraphics[width = \linewidth]{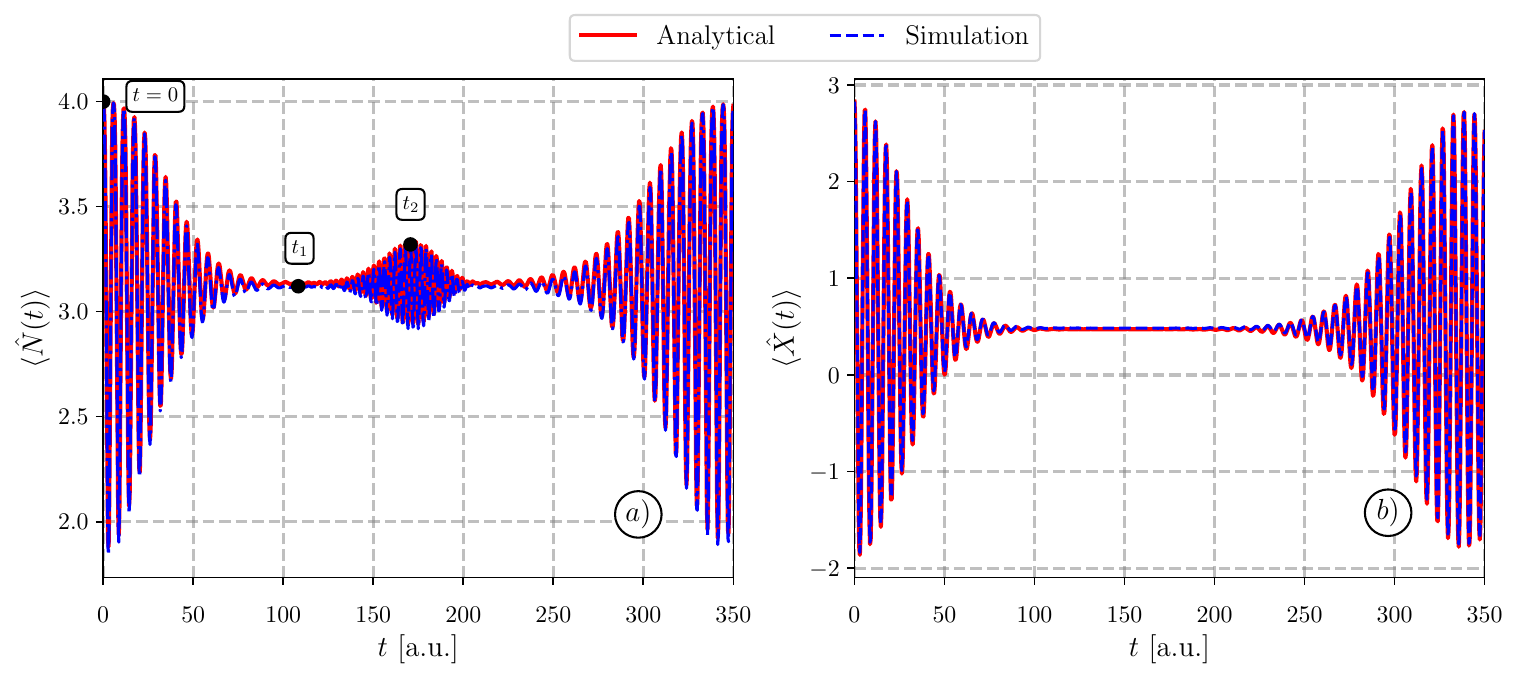}
    \caption{Comparison between analytical approximation and numerical simulation, with linear and quadratic coupling and an external field. Panel {\em a)} shows the mean expectation value of the phonon number operator and panel {\em b)} the average value of the quadrature, for an initial state $\ket{\psi(0)} = \ket{\alpha,\Gamma} = \ket{2,2}$. The Hamiltonian parameters are: $\omega_{c} = 10.0$, $\omega_{m} = 0.1\omega_{c}$, $g_{0} = 0.1$, $g_{1}=g_{0}^{2}$, $\Omega=0.25 \omega_{m}$ and $\omega_d=0.25 \omega_c$.}
    \label{fig:Nq_for}
\end{figure}
\begin{figure}[H]
    \centering
    \includegraphics[width = \linewidth]{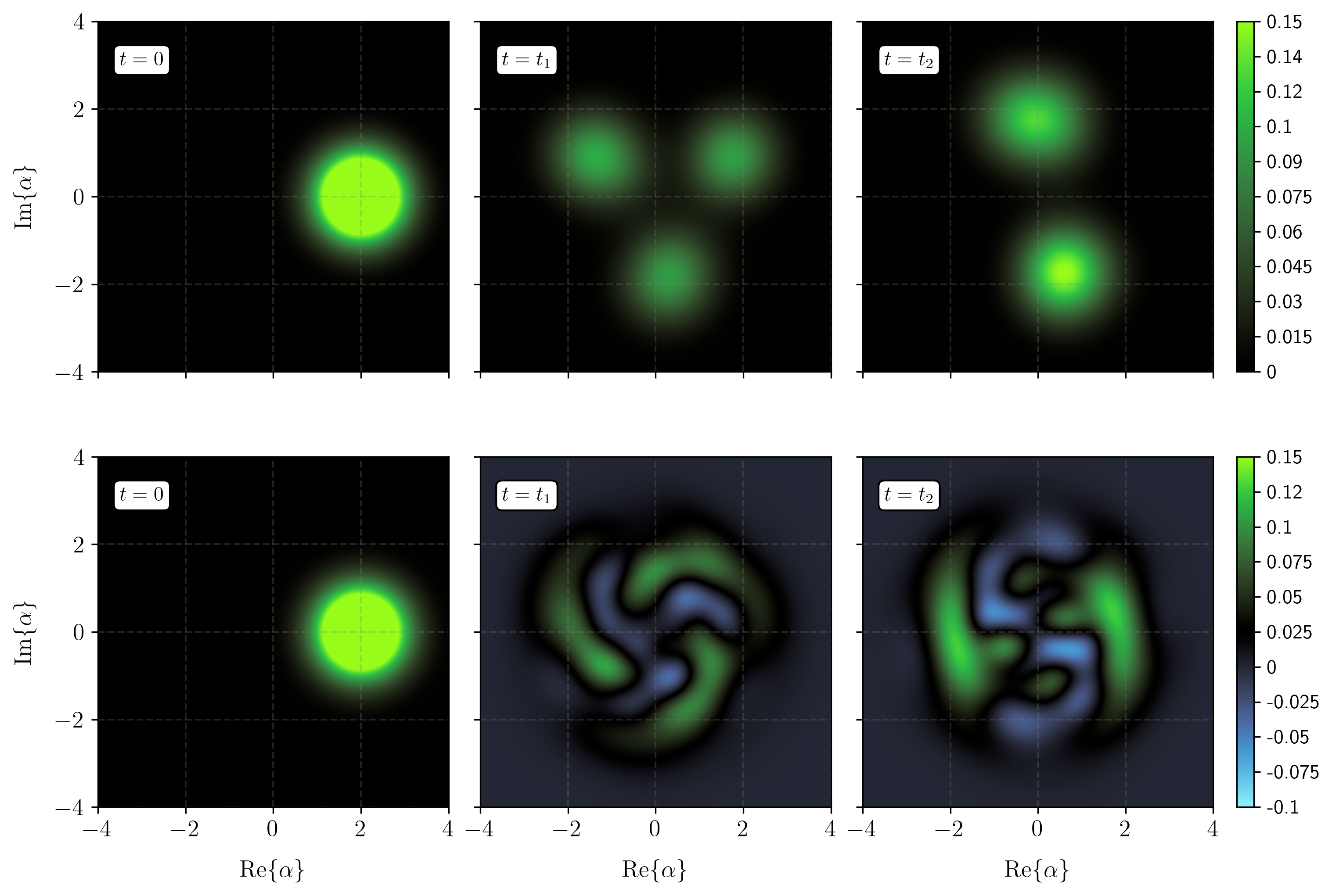}
    \caption{Wigner function evaluated at the times denoted in Fig.~\ref{fig:Nq_for}. The Hamiltonian parameters are: $\omega_{c} = 10.0$, $\omega_{m} = 0.1\omega_{c}$, $g_{0} = 0.1$, $g_{1}=g_{0}^{2}$, $\Omega=0.25 \omega_{m}$ and $\omega_d=0.25 \omega_c$. Above the Wigner function for the mechanical resonator and below that of the cavity field.}
    \label{Wigner_f}
\end{figure}
In figure \ref{fig:Nq_for} we show the results of our {\em{approximate}} analytical method and a numerical simulation for the temporal evolution of the average phonon number operator and the average of the quadrature position for an initial state given as a product of coherent states for the field and the mechanical oscillator. In full line (red) we show the results obtained with our approximation, and in broken line (blue) those obtained through a numerical simulation. In panel {\em{a)}} we show the temporal evolution of the average phonon number operator $\bar{N}(t)$. We can see a very  good agreement between the simulation and the approximate results, indicating that the approximation used for the photon number operator is justified, at least for the Hamiltonian parameters considered here. We stress the fact that we have chosen the forcing parameters such that the average photon number remains small. We have found a good agreement between our approximation and the numerical simulation whenever the forcing term is small, that is, when conditions $\omega_d\ll \omega_c$ and $\Omega<\omega_m$ are met. In  panel $a)$ we have marked times $t$ at the beginning of the evolution, $t_1$ at the center of a collapse, and $t_2$ at the middle of a revival since we will compute the Wigner function for the mechanical oscillator and the field at those specific instants of time. In panel {\em b)} we show the temporal evolution of the position quadrature, showing the presence of collapses and revivals. Both the calculations, the analytic and the numerical simulation yield practically the same results.

Finally, figure \ref{Wigner_f} shows in the first row, the Wigner function for the mechanical oscillator at the instants of time specified in figure \ref{fig:Nq_for}, and in the second row the Wigner function for the field at the same instants of time.  We see that at the initial time, both distributions correspond to a coherent state displaced from the origin in an amount $\Re{\alpha}=\Re{\Gamma} = 2$.
The Wigner function for the mechanical oscillator is positive at all the times we have considered. We see that at $t=t_1$, corresponding to the middle of a collapse, the distribution has been separated into three packets that do not interfere;  at $t_2$, in the middle of the first revival, we have two distinct packets that again do not interfere, in contrast with a Schr\"odinger cat where the different parts of the distribution show an interference pattern. The Wigner function for the field, at times $t_1$ and $t_2$, has been smeared in phase space and reaches negative values, indicating that the field is not a classical state. 


\section{Conclusions}
In this work, we have presented a Lie algebraic method to obtain the {\em exact} time evolution operator for an optomechanical system with linear and quadratic couplings. To illustrate the methodology, we first considered the case of a linear optomechanical system whose Hamiltonian contains an interaction term given as the product of the photon number operator and the mechanical oscillator displacement. We show that the set of operators in the Hamiltonian is closed under commutation; using the Wei-Norman theorem, we write the corresponding time evolution operator as a product of exponential functions. Then, we considered the case of a purely quadratic coupling, where the interaction term consists in the product of the photon number operator and the squared mechanical displacement. In this case, the operators in the Hamiltonian also form a closed Lie algebra, and we can write the time evolution operator in a product form. The case with linear and quadratic terms is dealt with, joining the two previous cases. A fundamental characteristic in these cases is that the photon number operator is a constant of the motion; thus it can be considered as a simple parameter and all the operators in the Hamiltonian belong to the mechanical oscillator. When we incorporate a forcing term linear in the creation-annihilation field operators, the photon number operator no longer commutes with the Hamiltonian, and it is not possible to obtain an exact expression for the time evolution operator. To consider a system with a low number of photons, we restricted the intensity of the driving, choosing a frequency far from resonance $\omega_d\ll \omega_c$ and a small forcing amplitude $\Omega <\omega_m$.  Under these conditions, we approximate the photon number operator in the interaction Hamiltonian by its time-dependent average. In so doing, we are left with an approximate interaction  with time-dependent coefficients whose only operators are those of the mechanical oscillator.   Using the time evolution operators obtained for the unforced and forced cases, we evaluated the evolution of several physical observables like the average phonon number operator, the position, and momentum quadratures and their dispersions and confronted these results against a numerical simulation taking the whole Hamiltonian.  We found an excellent agreement between them.   We evaluated the Mandel Q parameter temporal evolution for the mechanical oscillator states and found that, depending on the intensity of the linear coupling constant, its statistical properties can be super-Poissonian or sub-Poissonian whenever the quadratic coupling constant is not zero. When the quadratic coupling constant is equal to zero, the Mandel Q parameter is equal to one regardless of the value of the linear coupling, as corresponds to a coherent state.  Finally, we evaluated the Wigner function for the mechanical oscillator at some chosen times, and we found that it is always positive though its distribution splits into several distinct parts  while that of the field attain negative values at these same instants of time. 


\section*{Acknowledgements}
We acknowledge partial support from DGAPA-UNAM project IN109822. A.R.-U. acknowledges financial support by UNAM Posdoctoral Program (POSDOC) 2024-2025, and to ICF-UNAM for the assistance in-place. We also thank Reyes Garc\'ia (C\'omputo-ICF) for maintaining our computing servers.

\bibliography{bib}

\begin{thebibliography}{10}

\bibitem{Einstein:1909}
Anna Beck and Havas Havas, editors.
\newblock {\em The {Swiss} Years: Writings, 1900--1909. {English} Translation
  Supplement}, volume~2 of {\em The collected papers of Albert Einstein}.
\newblock Electronic, 1989.

\bibitem{Metcalf1999}
Harold~J. Metcalf and Peter van~der Straten.
\newblock {\em Laser Cooling and Trapping}.
\newblock Springer New York, 1999.

\bibitem{Wineland75}
DJ~Wineland and Hans Dehmelt.
\newblock Proposed $10^{14}\delta\nu/\nu$ laser fluorescence spectroscopy on
  {T}l+ mono-ion oscillator.
\newblock {\em Bull. Am. Phys. Soc}, 20(637), 1975.

\bibitem{Anderson1995}
M.~H. Anderson, J.~R. Ensher, M.~R. Matthews, C.~E. Wieman, and E.~A. Cornell.
\newblock Observation of {Bose-Einstein} condensation in a dilute atomic vapor.
\newblock {\em Science}, 269(5221):198–201, July 1995.

\bibitem{Monroe1996}
C.~Monroe, D.~M. Meekhof, B.~E. King, and D.~J. Wineland.
\newblock A “{S}chr\"{o}dinger cat” superposition state of an atom.
\newblock {\em Science}, 272(5265):1131–1136, May 1996.

\bibitem{Ketterle1995}
K.~B. Davis, M.~O. Mewes, M.~R. Andrews, N.~J. van Druten, D.~S. Durfee, D.~M.
  Kurn, and W.~Ketterle.
\newblock {Bose-Einstein} condensation in a gas of sodium atoms.
\newblock {\em Physical Review Letters}, 75(22):3969–3973, November 1995.

\bibitem{Ashkin78}
A.~Ashkin.
\newblock {Trapping of Atoms by Resonance Radiation Pressure}.
\newblock {\em Physical Review Letters}, 40(12):729–732, March 1978.

\bibitem{Mancini1994}
S.~Mancini and P.~Tombesi.
\newblock Quantum noise reduction by radiation pressure.
\newblock {\em Physical Review A}, 49(5):4055–4065, May 1994.

\bibitem{Jacobs1994}
K.~Jacobs, P.~Tombesi, M.~J. Collett, and D.~F. Walls.
\newblock Quantum-nondemolition measurement of photon number using radiation
  pressure.
\newblock {\em Physical Review A}, 49(3):1961–1966, March 1994.

\bibitem{Knight97}
S.~Bose, K.~Jacobs, and P.~L. Knight.
\newblock Preparation of nonclassical states in cavities with a moving mirror.
\newblock {\em Physical Review A}, 56(5):4175–4186, November 1997.

\bibitem{Manko97}
S.~Mancini, V.~I. Man’ko, and P.~Tombesi.
\newblock Ponderomotive control of quantum macroscopic coherence.
\newblock {\em Physical Review A}, 55(4):3042–3050, April 1997.

\bibitem{Thompson2008}
J.~D. Thompson, B.~M. Zwickl, A.~M. Jayich, Florian Marquardt, S.~M. Girvin,
  and J.~G.~E. Harris.
\newblock Strong dispersive coupling of a high-finesse cavity to a
  micromechanical membrane.
\newblock {\em Nature}, 452(7183):72–75, March 2008.

\bibitem{Favero2009}
Ivan Favero and Khaled Karrai.
\newblock Optomechanics of deformable optical cavities.
\newblock {\em Nature Photonics}, 3(4):201–205, April 2009.

\bibitem{Aspelmeyer14}
Markus Aspelmeyer, Tobias~J. Kippenberg, and Florian Marquardt.
\newblock Cavity optomechanics.
\newblock {\em Reviews of Modern Physics}, 86(4):1391–1452, December 2014.

\bibitem{Rai2008}
Amit Rai and G.~S. Agarwal.
\newblock Quantum optical spring.
\newblock {\em Physical Review A}, 78:013831, 7 2008.

\bibitem{Shi2013}
H.~Shi and M.~Bhattacharya.
\newblock Quantum mechanical study of a generic quadratically coupled
  optomechanical system.
\newblock {\em Physical Review A}, 87:043829, 4 2013.

\bibitem{Nunnenkamp2010}
A.~Nunnenkamp, K.~Børkje, J.~G.~E. Harris, and S.~M. Girvin.
\newblock Cooling and squeezing via quadratic optomechanical coupling.
\newblock {\em Physical Review A}, 82:021806, 8 2010.

\bibitem{jie14}
J.Q. Liao and F.~Nori.
\newblock Single-photon quadratic optomechanics.
\newblock {\em Scientific Reports}, 4:6302, 2014.

\bibitem{Xie2016}
Xie Hong, Lin Gong-Wei, Chen Xiang, Chen Zhi-Hua, and Lin Xiu-Min.
\newblock Single-photon nonlinearities in a strongly driven optomechanical
  system with quadratic coupling.
\newblock {\em Physical Review A}, 93:063860, 6 2016.

\bibitem{Kim2015}
Eun-Jong Kim, J.~R. Johansson, and Franco Nori.
\newblock Circuit analog of quadratic optomechanics.
\newblock {\em Physical Review A}, 91:033835, 3 2015.

\bibitem{Xu2018}
Xun-Wei Xu, Hai-Quan Shi, Ai-Xi Chen, and Yu-Xi Liu.
\newblock Cross-correlation between photons and phonons in quadratically
  coupled optomechanical systems.
\newblock {\em Physical Review A}, 98:013821, 7 2018.

\bibitem{Zhang2019}
Jian-Song Zhang, Ming-Cui Li, and Ai-Xi Chen.
\newblock Enhancing quadratic optomechanical coupling via a nonlinear medium
  and lasers.
\newblock {\em Physical Review A}, 99:013843, 1 2019.

\bibitem{machado2019}
J.~D.~P. Machado, R.~J. Slooter, and Ya.~M. Blanter.
\newblock Quantum signatures in quadratic optomechanics.
\newblock {\em Physical Review A}, 99:053801, 5 2019.

\bibitem{Kundu2021}
Akash Kundu, Chao Jin, and Jia-Xin Peng.
\newblock Study of the optical response and coherence of a quadratically
  coupled optomechanical system.
\newblock {\em Physica Scripta}, 96:065102, 6 2021.

\bibitem{Vanner2011}
Michael~R. Vanner.
\newblock {Selective Linear or Quadratic Optomechanical Coupling via
  Measurement}.
\newblock {\em Physical Review X}, 1:021011, 11 2011.

\bibitem{Satya2019}
U.~Satya Sainadh and M.~Anil Kumar.
\newblock Mimicking a hybrid-optomechanical system using an intrinsic quadratic
  coupling in conventional optomechanical system.
\newblock {\em Journal of Modern Optics}, 66:494--501, 3 2019.

\bibitem{Khorasani2017}
Sina Khorasani.
\newblock {Higher‐Order Interactions in Quantum Optomechanics: Revisiting
  Theoretical Foundations}.
\newblock {\em Applied Sciences}, 7:656, 6 2017.

\bibitem{Zhang2018}
X.~Y. Zhang, Y.~H. Zhou, Y.~Q. Guo, and X.~X. Yi.
\newblock Optomechanically induced transparency in optomechanics with both
  linear and quadratic coupling.
\newblock {\em Physical Review A}, 98:053802, 11 2018.

\bibitem{Gu2019}
Wen‐Ju Gu, Zhen Yi, Yan Yan, and Li‐Hui Sun.
\newblock {Generation of Optical and Mechanical Squeezing in the
  Linear‐and‐Quadratic Optomechanics}.
\newblock {\em Annalen der Physik}, 531, 8 2019.

\bibitem{Chao2021}
Shi-Lei Chao, Zhen Yang, Cheng-Song Zhao, Rui Peng, and Ling Zhou.
\newblock Force sensing in a dual-mode optomechanical system with
  linear–quadratic coupling and modulated photon hopping.
\newblock {\em Optics Letters}, 46:3075, 7 2021.

\bibitem{kippenberg}
A.~Schliesser, P.~DelHaye, N.~Nooshi, Vahala K.J., and T.~J. Kippenberg.
\newblock Radiation pressure cooling of a micromechanical oscillator using
  dynamical backaction.
\newblock {\em Phys. Rev. Lett.}, 97(24), 2006.

\bibitem{Meystre2013}
Pierre Meystre.
\newblock A short walk through quantum optomechanics.
\newblock {\em Annalen der Physik}, 525(3):215–233, December 2012.

\bibitem{purdy10}
T.P. Purdy, D.W.C. Brooks, T.~Botter, N.~Brahms, Z.-Y. Ma, and D.M.
  Stamper-Kurn.
\newblock Tunable cavity optomechanics with ultracold atoms.
\newblock {\em Physical Review Letters}, 105:133602, 2010.

\bibitem{recamier24}
L.~{Medina-Dozal}, J.~Récamier, H.~M. {Moya-Cessa}, F.~{Soto-Eguibar},
  R.~{Rom\'an-Ancheyta}, I.~{Ramos-Prieto}, and A.~R. Urzúa.
\newblock Temporal evolution of a driven optomechanical system in the strong
  coupling regime.
\newblock {\em Physica Scripta}, 99(1):015114, December 2023.

\bibitem{roman17}
R.~Rom\'an-Ancheyta, C.~Gonz\'alez-Guti\'errez, and J.~R\'ecamier.
\newblock Influence of the {K}err nonlinearity in a single nonstationary cavity
  mode.
\newblock {\em Journal of the Optical Society of America B}, 34(6):1170, 2017.

\bibitem{wei-norman}
James Wei and Edward Norman.
\newblock {Lie Algebraic Solution of Linear Differential Equations}.
\newblock {\em Journal of Mathematical Physics}, 4(4):575–581, April 1963.

\bibitem{recamier2011}
O~de~los {Santos-S\'anchez} and J~R\'ecamier.
\newblock Nonlinear coherent states for nonlinear systems.
\newblock {\em Journal of Physics A: Mathematical and Theoretical},
  44(14):145307, March 2011.

\bibitem{recamier-entropy}
L.~Medina-Dozal, I.~Ramos-Prieto, and J.~R\'ecamier.
\newblock Approximate evolution for a hybrid system: An optomechanical
  {J}aynes-{C}ummings model.
\newblock {\em Entropy}, 22:1373, 2020.

\bibitem{verhagen2012}
E.~Verhagen, S.~Deléglise, S.~Weis, A.~Schliesser, and T.~J. Kippenberg.
\newblock Quantum-coherent coupling of a mechanical oscillator to an optical
  cavity mode.
\newblock {\em Nature}, 482(7383):63–67, February 2012.

\bibitem{jayich08}
A.~M.~Jayich et. al.
\newblock Dispersive optomechanics: a membrane inside a cavity.
\newblock {\em New J. Phys.}, 10:095008, 2008.

\bibitem{sankey10}
J.~C. Sankey, C.~Yang, B.~M. Zwickl, A.~M. Jayich, and J.~G.~E.s Harris.
\newblock Strong and tunable nonlinear optomechanical coupling in a low-loss
  system.
\newblock {\em Nature Physics}, 6:707--712, 2010.

\bibitem{girvin10}
A.~Nunnenkamp, K.~Borje, J.~G.~E. Harris, and S.~M. Girvin.
\newblock Cooling and squeezing via quadratic optomechanical coupling.
\newblock {\em Phys. Rev. A}, 82:021806(R), 2010.

\bibitem{jahne09}
K.~Jahne, C.~Genes, K.~Hammerer, M.~Wallquist, E.~S. Polzik, and P.~Zoller.
\newblock Cavity-assisted squeezing of a mechanical oscillator.
\newblock {\em Prysical Review A}, 79:063819, 2009.

\bibitem{rabl11}
P.~Rabl.
\newblock Photon blockade effect in optomechanical systems.
\newblock {\em Physical Review Letters}, 107:063601, 2011.

\bibitem{nunnenkamp11}
A.~Nunnenkamp, K.~Borkje, and S.~M. Girvin.
\newblock Single photon optomechanics.
\newblock {\em Physical Review Letters}, 107:063602, 2011.

\end{thebibliography}
\bibliographystyle{unsrt}

\end{document}